\pdfoutput=1
\documentclass[iop]{emulateapj}

\slugcomment{}

\shorttitle{Multi-wavelength study of the star-formation in the S237 H\,{\sc ii} region}
\shortauthors{L.~K. Dewangan et al.}
\begin{document}

\title{Multi-wavelength study of the star-formation in the S237 H\,{\sc ii} region}
\author{L.~K. Dewangan\altaffilmark{1}, D.~K. Ojha\altaffilmark{2}, I. Zinchenko\altaffilmark{3}, P. Janardhan\altaffilmark{1}, and A. Luna\altaffilmark{4}}
\email{lokeshd@prl.res.in}
\altaffiltext{1}{Physical Research Laboratory, Navrangpura, Ahmedabad - 380 009, India.}
\altaffiltext{2}{Department of Astronomy and Astrophysics, Tata Institute of Fundamental Research, Homi Bhabha Road, Mumbai 400 005, India.}
\altaffiltext{3}{Institute of Applied Physics of the Russian Academy of Sciences, 46 Ulyanov st., Nizhny Novgorod 603950, Russia.}
\altaffiltext{4}{Instituto Nacional de Astrof\'{\i}sica, \'{O}ptica y Electr\'{o}nica, Luis Enrique Erro \# 1, Tonantzintla, Puebla, M\'{e}xico C.P. 72840.}
\begin{abstract}
We present a detailed multi-wavelength study of observations from X-ray, near-infrared to centimeter wavelengths 
to probe the star formation processes in the S237 region. 
Multi-wavelength images trace an almost sphere-like shell morphology of the region, which is filled with the 0.5--2 keV X-ray emission. 
The region contains two distinct environments - a bell-shaped cavity-like structure containing the peak of 1.4 GHz emission at center, 
and elongated filamentary features without any radio detection at edges of the sphere-like shell - where {\it Herschel} clumps are detected. 
Using the 1.4 GHz continuum and $^{12}$CO line data, the S237 region is found to be excited by a radio spectral type of B0.5V star and is associated 
with an expanding H{\sc ii} region. The photoionized gas appears to be responsible for the origin of the bell-shaped structure. 
The majority of molecular gas is distributed toward a massive {\it Herschel} clump 
(M$_{clump}$ $\sim$260 M$_{\odot}$), which contains the filamentary features and has a noticeable velocity gradient.
The photometric analysis traces the clusters of young stellar objects (YSOs) mainly toward the bell-shaped structure and the filamentary features.
Considering the lower dynamical age of the H\,{\sc ii} region (i.e. 0.2-0.8 Myr), these clusters are unlikely to be formed by the expansion of the H\,{\sc ii} region. 
Our results also show the existence of a cluster of YSOs and a massive clump at the intersection of filamentary features, 
indicating that the collisions of these features may have triggered cluster formation, similar to those found in Serpens South region. 
 \end{abstract}
\keywords{dust, extinction -- H\,{\sc ii} regions -- ISM: clouds -- ISM: individual objects (Sh 2-237) -- stars: formation -- stars: pre-main sequence} 
\section{Introduction}
\label{sec:intro}
Massive stars ($\gtrsim$ 8 M$_{\odot}$) produce a flood of ultraviolet (UV) photons, radiation pressure, 
and drive strong winds, which allow them to interact with the surrounding interstellar medium (ISM). 
In star-forming regions, the radio and infrared observations have revealed the ring/shell/bubble/filamentary features surrounding 
the H\,{\sc ii} regions associated with the OB stars, indirectly tracing the signatures of the energetics of powering sources \citep[e.g.][]{deharveng10,watson10,dewangan16}.
However, the physical processes of their interaction and feedback in their vicinity are still poorly understood \citep{zin07,tan14}.
In recent years, with the availability of {\it Herschel} observations, the study of initial conditions of cluster formation toward 
the filaments, in particular, has received much attention in star formation research. 
However, the role of filaments in the formation of dense massive star-forming clumps and clusters is still a matter of debate 
\citep[e.g.][]{myers09,schneider12,nakamura14,tan14}. 

Located at a distance of 2.3 kpc \citep{pandey13,lim15}, the star-forming region, Sh 2-237 
(hereafter S237; also known as NGC 1931/IRAS 05281+3412/RAFGL 5144/G173.9+0.3; $\alpha_{2000}$ = 05$^{h}$ 31$^{m}$ 22.8$^{s}$, 
$\delta_{2000}$ = +34$\degr$ 13$\arcmin$ 59$\arcsec$) has a broken or incomplete ring or shell-like appearance 
at wavelengths longer than 2 $\mu$m \citep[see Figure~1 in][]{pandey13}. 
One can find a comprehensive review of this star-forming region in \citet{reipurth08}. 
The H\,{\sc ii} region associated with the S237 region (hereafter S237 H\,{\sc ii} region) is ionized by a star of spectral type B0.5 \citep{glushkov75} or two B2 main-sequence-type stars \citep{pandey13}. 
\citet{balser11} estimated the velocity of ionized gas to be about $-$0.64 km s$^{-1}$ in the S237 H\,{\sc ii} region 
using a hydrogen radio recombination line (H87-93$\alpha$). 
Using CO (J = 1--0) line data, \citet{yang02} reported the radial velocity of molecular gas to be 
about $-$5.86 km s$^{-1}$ toward IRAS 05281+3412. \citet{leisawitz89} studied the molecular gas 
content of 34 open star clusters including S237 using $^{12}$CO (1-0) emission and suggested 
that a part of S237 is obscured by a molecular cloud. 
In the S237 region, \citet{pandey13} reported two separate stellar clusters and 
estimated a mean age of the young stellar objects (YSOs) to be 2$\pm$1 Myr. 
Based on the morphology of the region, ages of the YSOs and ionizing sources, 
they concluded a triggered star formation process in the S237 region. 
Using the isochrone fitting in the Hertzsprung-Russell diagram, \citet{lim15} reported 
a median age of 2 Myr with a spread of 4.5 Myr for the age distribution of the pre-main sequence (PMS) members in NGC 1931.

These previous studies suggest the presence of massive star(s) and ongoing star formation activity in the S237 region. 
The region is a relatively nearby star-forming site and has an interesting morphology, indicating it is a promising site to 
explore the impact of ionizing star(s). Despite the availability of different observational data-sets, the feedback of massive star(s) 
is not systematically studied in the S237 region. The physical conditions in the S237 region are yet to be 
determined and the velocity structure of molecular gas is also unknown. 
Recently, {\it Herschel} observations have revealed that filaments are often seen in the star-forming regions \citep[e.g.,][]{molinari10,andre10}. 
However, the identification of dense clumps and filaments is still lacking in the S237 region. Furthermore, the star formation toward 
clumps and filaments in S237 is yet to be probed. 
A knowledge of the physical environment of the region is very important to assess the ongoing physical mechanisms. 
In the present paper, we study in detail the physical processes responsible for the interaction and feedback effect of massive star(s) 
on its surroundings. To understand the ongoing physical mechanisms in S237, we examine the distribution of dust temperature, 
column density, extinction, ionized emission, hot gas, neutral hydrogen gas, kinematics of molecular gas, and YSOs using the multi-wavelength data. 

In Section~\ref{sec:obser}, we describe the multi-band data-sets used in the present work. 
In Section~\ref{sec:data}, we provide the details of results obtained. 
In Section~\ref{sec:disc}, we discuss the possible star formation scenario based on our findings. 
Finally, the results are summarized and concluded in Section~\ref{sec:conc}.
\section{Data and analysis}
\label{sec:obser}
In the present work, we selected a region of $\sim$30$\farcm6$ $\times$ 30$\farcm6$
(central coordinates: $l$ = 173$\degr$.993; $b$ = 0$\degr$.273) around the S237 region. 
In the following, a brief description of the multi-wavelength data spanning from X-ray, optical H$\alpha$, near-infrared (NIR) to radio wavelengths is presented. 

\subsection{X-ray data}
The ROSAT 0.5--2 keV X-ray image was obtained from the public archives 
(observation ID: WG931113P\_N1\_SI01.N1) maintained at the 
High Energy Astrophysics Science Archive Research Center (HEASARC) in USA. 
The pixel scale of the image is 45$\arcsec$. 
One can obtain more details about the ROSAT observations in \citet{voges99}.
A single ROSAT source \citep[i.e., J053134.4+341242;][]{voges00} is found in the S237 region. 
In the catalog, the total X-ray count was reported with an extraction radius of 300$\arcsec$ \citep{voges99}, 
indicating the presence of an extended diffuse X-ray emission.
\subsection{H$\alpha$ Narrow-band Image}
Narrow-band H$\alpha$ image at 0.6563 $\mu$m was obtained from the 
Isaac Newton Telescope Photometric H$\alpha$ Survey of the Northern Galactic Plane \citep[IPHAS;][]{drew05} survey database. 
The survey was made using the Wide-Field Camera (WFC) at the 2.5-m INT, located at La Palma. 
The WFC consists of four 4k $\times$ 2k CCDs, in an L-shape configuration. The pixel
scale is $0\farcs33$ and the instantaneous field is about 0.3 square degrees. 
One can find more details about the IPHAS survey in \citet{drew05}. 
\subsection{NIR (1--5 $\mu$m) Data}
NIR photometric {\it JHK} magnitudes of point sources have been obtained from the 
UKIDSS Galactic Plane Survey \citep[GPS;][]{lawrence07} sixth archival data release (UKIDSSDR6plus) and the Two Micron All Sky Survey \citep[2MASS;][]{skrutskie06}. 
Note that the UKIDSS GPS data are not available for the entire selected region and are available only for the zone-I area 
as highlighted by a dotted-line in Figure~\ref{fig1a}b. Hence, 2MASS data are employed for the area where UKIDSS GPS observations are absent.
The UKIDSS observations (resolution $\sim$$0\farcs8$) were taken using the UKIRT Wide Field Camera \citep[WFCAM;][]{casali07}. 
The UKIDSS GPS photometric data were calibrated using the 2MASS data. 
We obtained only reliable UKIDSS GPS photometric data, following the conditions listed in \citet{lucas08} and \citet{dewangan15}. 
2MASS data were also retrieved for bright sources that were saturated in the GPS catalog. 
To obtain reliable 2MASS photometric data, only those sources from the 2MASS catalog are chosen for the study that have photometric magnitude error of 0.1 and less in each band. 

Warm-{\it Spitzer} IRAC 3.6 and 4.5 $\mu$m photometric images (resolution $\sim$2$\arcsec$) and magnitudes of point sources have been retrieved 
from the Glimpse360\footnote[1]{http://www.astro.wisc.edu/sirtf/glimpse360/} \citep{whitney11} survey.
The photometric magnitudes were obtained from the GLIMPSE360 highly reliable catalog. To obtain further reliable Glimpse360 photometric data, only those sources are selected that have photometric magnitude error of 0.2 and less in each band. 
\subsection{Mid-infrared (12--22 $\mu$m) Data}
We utilized the publicly available archival WISE\footnote[2]{Wide Field Infrared Survey Explorer, which is a joint project of the
University of California and the JPL, Caltech, funded by the NASA} \citep{wright10} images 
at mid-infrared (MIR) 12 $\mu$m (spatial resolution $\sim$6$\arcsec$) and 22 $\mu$m (spatial resolution $\sim12\arcsec$). 
{\it WISE} photometric sensitivity\footnote[3]{http://wise2.ipac.caltech.edu/docs/release/allsky/} is reported to be 0.86 and 5.4 mJy (11.3 and 8.0 Vega mag) at 12 and 22 $\mu$m, respectively, in unconfused regions on the ecliptic plane. 
Saturation affects photometry for objects brighter than approximately 3.8 and -0.4 mag at 12 and 22 $\mu$m, respectively. 
\subsection{Far-infrared and Sub-millimeter Data}
\label{subsec:her}
Far-infrared (FIR) and sub-millimeter (sub-mm) continuum images were downloaded from the 
{\it Herschel} Space Observatory data archives. The processed level2$_{-}$5 images 
at 70 $\mu$m, 160 $\mu$m, 250 $\mu$m, 350 $\mu$m, and 500 $\mu$m were 
obtained using the {\it Herschel} Interactive Processing Environment \citep[HIPE,][]{ott10}. 
The beam sizes of the {\it Herschel} images at 70 $\mu$m, 160 $\mu$m, 250 $\mu$m, 350 $\mu$m, 
and 500 $\mu$m are 5$\farcs$8, 12$\arcsec$, 18$\arcsec$, 25$\arcsec$, and 37$\arcsec$, respectively \citep{poglitsch10,griffin10}. 
In this work, {\it Herschel} continuum images are utilized to compute the {\it Herschel} temperature and column density maps of the S237 region. 
Following the method described in \citet{mallick15}, we obtained the {\it Herschel} temperature and column density maps 
from a  pixel-by-pixel spectral energy distribution (SED) fit with a modified blackbody to the cold dust emission in 
the {\it Herschel} 160--500 $\mu$m wavelengths \citep[also see][]{dewangan15}. 
The {\it Herschel} 70 $\mu$m data are not used in the analysis, because the 70 $\mu$m emission is 
dominated by the UV-heated warm dust. 
In the following, a brief step-by-step description of the procedures is described. 

The {\it Herschel} 160 $\mu$m 
image is calibrated in the unit of Jy pixel$^{-1}$, 
while the images at 250--500 $\mu$m are calibrated in the surface brightness unit of MJy sr$^{-1}$. 
The plate scales of the 160, 250, 350, and 500 $\mu$m images are 6.4, 6, 10, and 14 arcsec/pixel, respectively.  
In the first step, prior to the SED fit, the 160, 250, and 350 $\mu$m images were convolved to the lowest angular 
resolution of 500 $\mu$m image ($\sim$37$\arcsec$) 
and were converted into the same flux unit (i.e. Jy pixel$^{-1}$). 
Furthermore, these images were regridded to the pixel size of 500 $\mu$m image ($\sim$14$\arcsec$). 
These steps were carried out using the convolution kernels available in the HIPE software. 
Next, the sky background flux level was determined to be 0.051, 0.115, 0.182, and $-$0.0009 Jy pixel$^{-1}$ for the 500, 350, 250, and 
160 $\mu$m images (size of the selected region $\sim$5$\farcm$6 $\times$ 7$\farcm$8; 
centered at:  $l$ = 174$\degr$.127; $b$ = $-$0$\degr$.222), respectively. 
To avoid diffuse emission associated with the selected target, the featureless dark area away from the S237 region was carefully chosen 
for the background estimation. 

Finally, to obtain the maps, we fitted the observed flux densities at each pixel with the modified blackbody 
model \citep[see equations 8 and 9 given in][]{mallick15}. 
The fitting was done using the four data points for each pixel, retaining the 
dust temperature (T$_{d}$) and the column density ($N(\mathrm H_2)$) 
as free parameters. 
In the analysis, we adopted a mean molecular weight per hydrogen molecule ($\mu_{H2}$=) 2.8 
\citep{kauffmann08} and an absorption coefficient ($\kappa_\nu$ =) 0.1~$(\nu/1000~{\rm GHz})^{\beta}$ cm$^{2}$ g$^{-1}$, 
including a gas-to-dust ratio ($R_t$ =) of 100, with a dust spectral index of $\beta$\,=\,2 \citep[see][]{hildebrand83}. 
The final {\it Herschel} temperature and column density maps are presented in Section~\ref{subsec:temp}.
\subsection{Molecular CO line data}
The observations of $^{12}$CO(1-0) and $^{13}$CO(1-0) emission were taken using the Five College Radio Astronomy 
Observatory (FCRAO) 14 meter telescope in New Salem, Massachusetts. 
The FCRAO beam sizes are 45$\arcsec$ (with angular sampling of 22$\farcs$5) and 46$\arcsec$ (with angular sampling of 22$\farcs$5) 
for $^{12}$CO and $^{13}$CO, respectively. Each line data have a velocity resolution of 0.25~km\,s$^{-1}$. 
Typical rms values for the spectra are 0.25 K for $^{12}$CO and 0.2 K for $^{13}$CO \citep[e.g.][]{heyer96}.
The S237 region was observed as part of the Extended Outer Galaxy Survey \citep[E-OGS,][]{brunt04}, 
that extends the coverage of the FCRAO Outer Galaxy Survey \citep[OGS,][]{heyer98} to Galactic longitude ($l$) = 193$\degr$, 
over a Galactic latitude ($b$) range of $-$3$\degr$.5 $\leq$ $b$ $\leq$ +5$\degr$.5. 
These $^{12}$CO and $^{13}$CO data cubes were provided by M. Heyer and C. Brunt (through private communication).  
The FCRAO $^{12}$CO profile along the line of sight to the S237 region shows a single velocity component at v = $-$4.6 km s$^{-1}$.
\subsection{Radio continuum data}
Radio continuum map at 1.4 GHz was retrieved from the NRAO VLA Sky Survey (NVSS) archive. 
The survey covers the sky north of $\delta_{2000}$ = $-$40$\degr$ at 1.4 GHz with a beam of 45$\arcsec$ 
and a nearly uniform sensitivity of $\sim$0.45 mJy/beam \citep{condon98}.
\subsection{H\,{\sc i} line data}
21 cm H\,{\sc i} line data were obtained from the Canadian Galactic Plane Survey \citep[CGPS;][]{taylor03}.
The velocity resolution of {H\,{\sc i} line data is 1.32~km\,s$^{-1}$, sampled every
0.82~km\,s$^{-1}$. The data have a spatial resolution of 1$\arcmin$ $\times$ 1$\arcmin$ csc$\delta$.
The line data have a brightness-temperature sensitivity of $\Delta$T$_{B}$ = 3.5 sin$\delta$ K.
One can find more details about the CGPS observing and data processing strategy in \citet{taylor03}.
\section{Results}
\label{sec:data}
\subsection{Multi-band picture of S237}
\subsubsection{Continuum emission and gas distribution in S237}
\label{subsec:pic}
The longer wavelength data enable to penetrate deeper into the star-forming cloud despite the high extinction, allowing to infer the physical environment of a given star-forming region. In this section, we employ multi-band data to trace the physical structure of the S237 region.  
Figure~\ref{fig1a}a shows a two-color composite image using {\it Herschel} 70 $\mu$m in red and {\it WISE} 12.0 $\mu$m in green. 
A three-color composite image (350 $\mu$m (red), 250 $\mu$m (green), and 22 $\mu$m (blue)) is shown in Figure~\ref{fig1a}b.
The {\it Herschel} and {\it WISE} images depict a broken or incomplete ring or shell-like appearance of the S237 region at wavelengths longer than 2 $\mu$m. The images reveal a prominent bell-shaped cavity-like morphology and elongated filamentary features at the center 
and the edge of the shell-like structure, respectively (see Figures~\ref{fig1a}a and~\ref{fig1a}b). 
The inset on the bottom left shows the zoomed-in view around the IRAS 05281+3412 using the {\it Spitzer} 3.6 $\mu$m image (see Figure~\ref{fig1a}a). 
The filamentary features are also highlighted in the image. Figure~\ref{fig2a} shows the radio continuum and molecular emissions overlaid on the {\it Herschel} image at 500 $\mu$m. 
The bulk of the 160--500 $\mu$m emission comes from cold dust components (see Section~\ref{subsec:temp} for quantitative estimates), 
while the emission at 22 $\mu$m traces the warm dust emission (see Figure~\ref{fig1a}b). 
In Figure~\ref{fig2a}a, we show the spatial distribution of ionized emission traced in the NVSS 1.4 GHz map, concentrating near the bell-shaped cavity-like morphology. This also implies that the ionized emission is located within the shell-like structure. 
The radio continuum emission is absent toward the elongated filamentary features. 
In Figure~\ref{fig2a}b, we present the molecular $^{12}$CO (J = 1--0) gas emission in the direction 
of the S237 region, revealing at least four molecular condensations (designated as conds1-4). 
The $^{12}$CO profile traces the region in a velocity range between $-$7 and $-$2 km s$^{-1}$. 
The molecular condensations are also found toward the filamentary features and the bell-shaped cavity-like morphology. 
It can be seen that the majority of molecular gas is distributed toward the molecular condensation, conds1 where filamentary features are observed. 
The details of integrated $^{12}$CO map as well as kinematics of molecular gas are described in Section~\ref{sec:coem}.

Together, Figures~\ref{fig1a} and~\ref{fig2a} have allowed us to obtain a pictorial multi-wavelength view of the environment in the S237 region.
\subsubsection{Lyman continuum flux}
\label{subsec:radio}
In this section, we estimate the number of Lyman continuum photons using the radio continuum map 
which will allow to derive the spectral type of the powering candidate associated with the S237 H\,{\sc ii} region. 
In Figure~\ref{fig2a}a, the NVSS radio map traces a spherical morphology of the S237 H\,{\sc ii} region. 
We used the {\it clumpfind} IDL program \citep{williams94} to estimate the integrated flux density and 
the radio continuum flux was integrated up to 0.3\% contour level of peak intensity. 
Using the 1.4 GHz map, we computed the integrated flux density (S$_{\nu}$) and radius (R$_{HII}$) of the H\,{\sc ii} region to be 708 mJy and 1.27 pc, respectively. 
The integrated flux density allows us to compute the number of Lyman continuum photons (N$_{uv}$), using the following equation \citep{matsakis76}:
\begin{equation}
N_{uv} (s^{-1}) = 7.5\, \times\, 10^{46}\, \left(\frac{S_{\nu}}{Jy}\right)\left(\frac{D}{kpc}\right)^{2} 
\left(\frac{T_{e}}{10^{4}K}\right)^{-0.45} \\ \times\,\left(\frac{\nu}{GHz}\right)^{0.1}
\end{equation}
\noindent where S$_{\nu}$ is the measured total flux density in Jy, D is the distance in kpc, 
T$_{e}$ is the electron temperature, and $\nu$ is the frequency in GHz. 
This analysis is performed for the electron temperature of 10000~K and a distance of 2.3 kpc.
We obtain N$_{uv}$ (or logN$_{uv}$) to be $\sim$2.9 $\times$ 10$^{47}$ s$^{-1}$ (47.47) for the S237 H\,{\sc ii} region, 
which corresponds to a single ionizing star of spectral type B0.5V-B0V (see Table II in \citet{panagia73} for a theoretical value).  

The knowledge of N$_{uv}$ and R$_{HII}$ values is also used to infer the dynamical age (t$_{dyn}$) of the S237 H\,{\sc ii} region.
The age of the H\,{\sc ii} region can be computed at a given radius R$_{HII}$, using the following equation \citep{dyson80}:
\begin{equation}
t_{dyn} = \left(\frac{4\,R_{s}}{7\,c_{s}}\right) \,\left[\left(\frac{R_{HII}}{R_{s}}\right)^{7/4}- 1\right] 
\end{equation}
where c$_{s}$ is the isothermal sound velocity in the ionized gas (c$_{s}$ = 11 km s$^{-1}$; \citet{bisbas09}), 
R$_{HII}$ is previously defined, and R$_{s}$ is the radius of the Str\"{o}mgren sphere (= (3 N$_{uv}$/4$\pi n^2_{\rm{0}} \alpha_{B}$)$^{1/3}$, where 
the radiative recombination coefficient $\alpha_{B}$ =  2.6 $\times$ 10$^{-13}$ (10$^{4}$ K/T)$^{0.7}$ cm$^{3}$ s$^{-1}$ \citep{kwan97}, 
N$_{uv}$ is defined earlier, and ``n$_{0}$'' is the initial particle number density of the ambient neutral gas. 
Assuming typical value of n$_{0}$ (as 10$^{3}$(10$^{4}$) cm$^{-3}$), we calculated the 
dynamical age of the S237 H\,{\sc ii} region to be $\sim$0.2(0.8) Myr. 
\subsection{IRAC ratio map and H\,{\sc i} gas }
\label{subsec:h2out}
To infer the signatures of molecular outflows and the impact of massive stars on their surroundings, 
the {\it Spitzer}-IRAC ratio maps have been employed in combination with the radio continuum emission \citep{povich07,watson08,dewangan11,dewangan12,dewangan16}. 
Due to almost identical point response function (PRF) of IRAC 3.6 $\mu$m and 4.5 $\mu$m images, 
a ratio map of 4.5 $\mu$m/3.6 $\mu$m emission can be obtained directly using the ratio of 4.5 $\mu$m to 3.6 $\mu$m images.
IRAC 4.5 $\mu$m band harbors a hydrogen recombination line Br$\alpha$ (4.05 $\mu$m) and 
a prominent molecular hydrogen line emission ($\nu$ = 0--0 $S$(9); 4.693 $\mu$m), which can be excited by outflow shocks.
IRAC 3.6 $\mu$m band contains polycyclic aromatic hydrocarbon (PAH) emission at 3.3 $\mu$m as well as a prominent molecular hydrogen feature at 3.234 
$\mu$m ($\nu$ = 1--0 $ O$(5)). The ratio map of 4.5 $\mu$m/3.6 $\mu$m emission is shown in Figure~\ref{fig3a}a, tracing the bright and dark/black regions.
The shell-like morphology is depicted and the map traces the edges of the shell (see dark/black regions in Figure~\ref{fig3a}a). 
The 0.5--2 keV X-ray emission is detected within the shell-like morphology, indicating the presence of hot gas emission in the region (see Figure~\ref{fig3a}a). 
The prominent bell-shaped cavity and the elongated filamentary features are also seen in the ratio map 
(see arrows and highlighted dotted-dashed box in Figure~\ref{fig3a}a). 
In Figure~\ref{fig3a}b, the distribution of molecular gas, ionized gas, and hot gas emission is shown together. 
In ratio 4.5 $\mu$m/3.6 $\mu$m map, the bright emission regions suggest the domination of 4.5 $\mu$m emission, 
while the black or dark gray regions indicate the excess of 3.6 $\mu$m emission.  
In Figure~\ref{fig4af}, we present H$\alpha$ image overlaid with the NVSS 1.4 GHz and {\it Herschel} 70 $\mu$m emission. 
The H$\alpha$ image shows the extended diffuse H$\alpha$ emission in the S237 region which is well distributed within the shell-like morphology. 
Figure~\ref{fig4af} also shows the spatial match between the H$\alpha$ emission and the radio continuum emission.  
Figures~\ref{fig4a}a and~\ref{fig4a}b show zoomed-in ratio map and color-composite image toward the radio peak. 
A bell-shaped cavity is evident and the peaks of radio continuum 1.4 GHz emission and diffuse H$\alpha$ emission
are present within it (see Figures~\ref{fig4af} and~\ref{fig4a}b). 
The warm dust emission traced in the 70 $\mu$m image depicts the walls of the bell-shaped cavity (see Figure~\ref{fig4a}b). 
However, the peak of molecular $^{12}$CO(1-0) emission does not coincide with the radio continuum peak (see Figure~\ref{fig4a}). 
This indicates that the bell-shaped cavity can be originated due to the impact of ionized emission (also see Section~\ref{sec:feedb}).
In the ratio map, there are bright emission regions near the radio continuum emission, indicating the excess of 4.5 $\mu$m emission.
As mentioned above, the 4.5 $\mu$m band contains Br$\alpha$ feature at 4.05 $\mu$m. 
Hence, due to the presence of ionized emission, these bright emission regions probably trace the Br$\alpha$ 
features. In Figure~\ref{fig4a}a, the excess of 4.5 $\mu$m emission is also observed to the bottom of the bell-shaped morphology, 
which is designated as subreg1 and is coincident with the 0.5--2 keV X-ray peak emission. 
We also find bright emission region near the elongated filamentary features where the radio continuum emission is absent 
(see arrows in Figure~\ref{fig3a}a). Hence, this emission is probably tracing the outflow activities. 
Considering the presence of 3.3 $\mu$m PAH feature in the 3.6 $\mu$m band, 
the edges of the shell-like morphology around the H\,{\sc ii} region appear to depict 
photodissociation regions (or photon-dominated regions, or PDRs).

In Figure~\ref{fig4bb}, we present 21 cm H\,{\sc i} velocity channel maps of the S237 region. 
The bright H\,{\sc i} feature is seen near the 1.4 GHz emission, however the sphere-like shell morphology is depicted in black or dark gray regions.
It appears that the black or dark gray regions in the H\,{\sc i} channel maps trace the H\,{\sc i} self-absorption (HISA) features (i.e. shell-like HISA features) \citep[e.g.,][]{kerton05}. 
It has been suggested that the HISA features are produced by the residual amounts of very cold H\,{\sc i} gas in molecular clouds 
\citep{burton78,baker79, burton81,liszt81}. 
In the S237 region, the HISA features are more prominent in a velocity range of $-$2.29 to $-$3.11 km s$^{-1}$. 
The 21 cm H\,{\sc i} line data are presented here only for a morphological comparison with the infrared images.  
The 0.5--2 keV X-ray emission is also shown in a channel map (at $-$2.29 km s$^{-1}$), 
allowing to infer the distribution of cold H\,{\sc i} gas and hot gas emission in the region. 
In Figure~\ref{fig4uu}, to compare the morphology of the S237 region, we have shown infrared images 
(at 22 $\mu$m and 250 $\mu$m) and H\,{\sc i} map (at 21 cm). 
 We find that the shell-like HISA feature is spatial correlated with dust emissions as traced 
in the {\it Spitzer}, {\it WISE}, and {\it Herschel} images. 
The presence of H\,{\sc i} further indicates the PDRs surrounding the H\,{\sc ii} region and hot gas emission. 
\subsection{{\it Herschel} temperature and column density maps}
\label{subsec:temp}
The {\it Herschel} temperature and column density maps are used to infer the physical conditions present 
in a given star-forming region. 
In Figure~\ref{fig5a}, we show the final temperature and column density maps (resolution $\sim$37$\arcsec$). 
The procedures for calculating the {\it Herschel} temperature and column density maps were mentioned in 
Section~\ref{subsec:her}.

In the {\it Herschel} temperature map, the area near the H\,{\sc ii} region is traced with 
the considerably warmer gas (T$_{d}$ $\sim$29-47 K) (see Figure~\ref{fig5a}a). 
The {\it Herschel} temperature map traces the edges of the shell in a temperature range of about 22--28~K. 
Several condensations are found in the column density map and one of the condensations (see clump 1) has the highest column density 
(peak $N(\mathrm H_2)$ $\sim$3.6~$\times$~10$^{21}$ cm$^{-2}$; A$_{V}$ $\sim$3.8 mag) located toward the filamentary features (see Figure~\ref{fig5a}b). 
The relation between optical extinction and hydrogen column density 
\citep[$A_V=1.07 \times 10^{-21}~N(\mathrm H_2)$;][]{bohlin78} is used here. \citet{krumholz08} proposed a threshold value 
of 1 gm cm$^{-2}$ (or corresponding column densities $\sim$3 $\times$ 10$^{23}$ cm$^{-2}$) for formation of massive stars. 
This implies that the formation of massive stars in the identified {\it Herschel} clumps is unlikely.
The column density structure of the S237 appears non-homogeneous with higher column density clumps 
engulfed in a medium with lower column density. The bell-shaped cavity is also seen in the column density map. 

In the column density map, we employed the {\it clumpfind} algorithm to identify the clumps and their total column densities. 
We find 13 clumps, which are labeled in Figure~\ref{fig5a}b and their boundaries are also shown in Figure~\ref{fig5a}c.
Among 13 clumps, seven clumps (e.g., 1, 3, 4, 6, 7, 9, and 13) are found toward the edges of the shell-like structure, while other three clumps (e.g., 2, 5, and 11) are located within the shell-like structure and remaining three clumps (e.g., 8, 10, and 12) are away from the shell-like structure. 
To assess the spatial distribution of clumps with respect to the HISA features, these clumps are also marked in the CGPS 21 cm H\,{\sc i} single-channel map at $-$3.94 km s$^{-1}$ (see Figure~\ref{fig4bb}). 
The mass of each clump is computed using its total column density and can be determined using the formula:
\begin{equation}
M_{clump} = \mu_{H_2} m_H Area_{pix} \Sigma N(H_2)
\end{equation}
where $\mu_{H_2}$ is assumed to be 2.8, $Area_{pix}$ is the area subtended by one pixel, and 
$\Sigma N(\mathrm H_2)$ is the total column density. 
The mass of each {\it Herschel} clump is tabulated in Table~\ref{tab1}. 
The table also contains an effective radius of each clump, which is provided by the {\it clumpfind} algorithm. 
The clump masses vary between 10 M$_{\odot}$ and 260 M$_{\odot}$. 
The most massive clump (i.e. clump1) associated with the molecular condensation, conds1 (see Figure~\ref{fig2a}b) 
is away from the radio peak, while the clump2 linked with the molecular condensation, conds2 (see Figure~\ref{fig2a}b) 
is associated with the H\,{\sc ii} region. 
\subsection{Kinematics of molecular gas}
\label{sec:coem} 
In this section, we present distributions of $^{12}$CO (J=1--0) and $^{13}$CO (J=1--0) gas in the S237 region.
An inspection of the $^{12}$CO and $^{13}$CO line profiles reveals that the molecular cloud associated with the S237 region (i.e S237 molecular cloud) 
is well traced in a velocity range of $-$7 to $-$2 km s$^{-1}$. 
In general, $^{12}$CO emission is optically-thicker than $^{13}$CO. 
To study the spatial distribution of the molecular gas in the S237 region, in Figure~\ref{fig6a}, 
we present $^{12}$CO (J=1--0) velocity channel maps 
at different velocities within the velocity range from $-$6.25 to $-$1.75 km s$^{-1}$ in steps of 0.5 km s$^{-1}$.
The channel maps show at least four molecular condensations (i.e. conds1--4) and the bulk of the molecular gas is found toward the 
condensation ``conds1". In Figure~\ref{fig7a}, we show the $^{13}$CO (J=1--0) velocity channel maps as zoomed-in view toward the condensation ``conds1", where
 the high- and low-velocity gas distribution is evident toward the {\it Herschel} clump1. 
Note that the molecular condensation, ``conds1" contains the elongated filamentary features and a massive {\it Herschel} clump (i.e. clump1).  
In Figure~\ref{fig8a}, we present the integrated $^{12}$CO and $^{13}$CO intensity maps 
and the galactic position-velocity maps. In the position-velocity diagrams of the $^{12}$CO emission, 
we find a noticeable velocity gradient along the condensation ``conds1" (see Figures~\ref{fig8a}c and~\ref{fig8a}e) and an almost inverted C-like 
structure (see Figure~\ref{fig8a}e). 
In Figure~\ref{fig8a} (see right panels), we also show the integrated $^{13}$CO (1--0) intensity map and the position-velocity maps. 
The optical depth in $^{13}$CO is much lower than that in $^{12}$CO, and the $^{13}$CO data can trace dense region (n(H$_{2}$) $>$ 10$^{3}$ cm$^{-3}$). 
The integrated $^{13}$CO intensity map detects only the condensation ``conds1" which contains the {\it Herschel} clump1 (see Figure~\ref{fig5a}b). 
It suggests that the conds1 is the densest molecular condensations compared to other three condensations (i.e. conds2--4). 
Using the integrated $^{13}$CO intensity map (see Figure~\ref{fig8a}b), we compute the mass of the molecular condensation ``conds1" to be about 242 M$_{\odot}$, which is in 
agreement with the clump mass estimated using the {\it Herschel} data. 
In the calculation, we use an excitation temperature of 20 K, the abundance ratio (N(H$_{2}$)/N($^{13}$CO)) of 7 $\times$ 10$^{5}$, 
and the ratio of gas to hydrogen by mass of about 1.36. 
One can find more details about the clump mass estimation in \citet{yan16} \citep[see equations 4 and 5 in][]{yan16}.
The position-velocity diagrams of the $^{13}$CO emission clearly indicate the presence of a velocity 
gradient along the condensation ``conds1" (see Figures~\ref{fig8a}d and~\ref{fig8a}f). 
The condensation ``conds1" hosts a cluster of young populations (see Section~\ref{subsec:surfden}) and filamentary features. 
Therefore, one can suspect the presence of molecular outflows in the condensation ``conds1" (also see channel maps in Figure~\ref{fig7a}). 
The velocity gradient could also be explained as gas flowing through the filamentary features into their intersection. 
Due to the coarse beam of the FCRAO $^{12}$CO and $^{13}$CO data (beam size $\sim$45$\arcsec$), we do not further explore this aspect in this work 
and are also unable to infer the gas distribution toward the filamentary features. 

Recently, based on the molecular line data analysis and modeling work, \citet{arce11} suggested an inverted C-like or 
ring-like structure for an expanding shell/bubble in the position-velocity diagrams (see Figure~5 in \citet{arce11}). 
In the S237 region, the presence of an inverted C-like structure in the position-velocity diagram can indicate an expanding shell.
Additionally, as previously mentioned, the S237 H\,{\sc ii} region is excited by a single source of radio spectral B0.5V.
Hence, it appears that the S237 region is associated with an expanding H\,{\sc ii} region with 
an expansion velocity of the gas to be $\sim$1.65 km s$^{-1}$. 

All together, the FCRAO $^{12}$CO and $^{13}$CO data suggest the presence of the expanding H\,{\sc ii} region. 
\subsection{Young stellar objects in S237}
\subsubsection{Identification of young stellar objects}
\label{subsec:phot1}
The GLIMPSE360, UKIDSS-GPS, and 2MASS data allow us to investigate the infrared excess sources present in the S237 region. 
In the following, we describe the YSOs identification and classification schemes.\\

1. To identify infrared-excess sources, \citet{gutermuth09} described various conditions 
using the H, K, 3.6, and 4.5 $\mu$m data and utilized the dereddened color-color space ([K$-$[3.6]]$_{0}$ and [[3.6]$-$[4.5]]$_{0}$). 
Using the GLIMPSE360, UKIDSS-GPS, and 2MASS catalog, these dereddened colors were determined 
using the color excess ratios given in \citet{flaherty07}. 
This dereddened color-color space is also used to identify possible dim extragalactic contaminants from YSOs with additional 
conditions (i.e., [3.6]$_{0}$ $<$ 15 mag for Class~I and [3.6]$_{0}$ $<$ 14.5 mag for Class~II). 
The observed color and the reddening laws \citep[from][]{flaherty07} are utilized to compute the dereddened 3.6 $\mu$m magnitudes. 
This scheme yields 80 (7 Class~I and 73 Class~II) YSOs (see Figure~\ref{fig9a}a). \\

2. We find that some sources have detections only in the H and K bands. 
To further select infrared excess sources from these selected population, we utilized a 
color-magnitude (H$-$K/K) diagram (see Figure~\ref{fig9a}b). 
The diagram depicts the red sources with H$-$K $>$ 0.65 mag. This color criterion is selected based on the 
color-magnitude analysis of the nearby control field. 
We identify 18 additional YSO candidates using this scheme in our selected region.\\

Using the UKIDSS, 2MASS, and GLIMPSE360 data, a total of 98 YSOs are obtained in the selected region. 
The positions of all YSOs are shown in Figure~\ref{fig10a}a. 
\subsubsection{Spatial distribution of YSOs}
\label{subsec:surfden}
Using the nearest-neighbour (NN) technique, the surface density analysis of YSOs is a popular method to 
examine their spatial distribution in a given star-forming region 
\citep[e.g.][]{gutermuth09,bressert10,dewangan15}, which can be used to find the young stellar clusters. 
Using the NN method, we obtain the surface density map of YSOs in a manner similar to that described in \citet{dewangan15} 
\citep[also see equation in][]{dewangan15}. 
The surface density map of all the selected 98 YSOs was constructed, using a 5$\arcsec$ grid and 6 NN at a distance of 2.3 kpc. 
In Figure~\ref{fig10a}b, the surface density contours of YSOs are presented and are drawn at 3, 5, 7, 10, 15, and 20 YSOs/pc$^{2}$, increasing from the outer to the inner regions. 
In Figure~\ref{fig10a}b, two clusters of YSOs are observed in the S237 region and 
are mainly seen toward the {\it Herschel} clump1, clump2, and subreg1 (see Figure~\ref{fig10a}b). 
However, one cluster of YSOs seems to be linked with the {\it Herschel} clump2 and the subreg1 together.  
In Section~\ref{subsec:temp}, we have seen that the filamentary features and bell-shaped cavity are observed toward the 
{\it Herschel} clump1 and clump2, respectively. 
Hence, in the S237 region, the star formation activities are found toward the filamentary features, bell-shaped cavity, and subreg1 (see Figure~\ref{fig10a}b).
Recently, \citet{pandey13} also identified young populations in the S237 region using the optical and NIR (1--5 $\mu$m) data. 
Based on the spatial distribution of young populations, they found most of the YSOs distributed mainly toward the 
bell-shaped cavity, subreg1, and filamentary features \citep[see Figure 15 in][]{pandey13}, which was also reported by \citet{lim15} 
\citep[see Figure 8 in][]{lim15}. Furthermore, \citet{lim15} found an elongated shape of the surface density contours of PMS sources 
which is very similar in morphology to that investigated in this work. Taken together, these previous results are in a good agreement with our presented results.

In the molecular condensation, conds1, the filamentary features are physically associated with a cluster of YSOs, a massive clump (having the highest column density, i.e. 3.6 $\times$ 10$^{21}$ cm$^{-2}$), and molecular gas (see Figure~\ref{fig11a}). 
Interestingly, this result suggests the role of filaments in the star formation process. 
\subsection{feedback of a massive star}
\label{sec:feedb} 
Based on the radio continuum data analysis, we find that the S237 region is powered by a radio spectral type of B0.5V star.
To study the feedback of a massive star in its vicinity, we compute the three pressure components 
(i.e. pressure of an H\,{\sc ii} region $(P_{HII})$, radiation pressure (P$_{rad}$), and stellar wind ram pressure (P$_{wind}$)) 
driven by a massive star. 
These pressure components ($P_{HII}$, P$_{rad}$, and P$_{wind}$) are defined below \citep[e.g.][]{bressert12}:
\begin{equation}
P_{HII} = \mu m_{H} c_{s}^2\, \left(\sqrt{3N_{uv}\over 4\pi\,\alpha_{B}\, D_{s}^3}\right);\\ 
\end{equation}
\begin{equation}
P_{rad} = L_{bol}/ 4\pi c D_{s}^2; \\ 
\end{equation}
\begin{equation}
P_{wind} = \dot{M}_{w} V_{w} / 4 \pi D_{s}^2; \\
\end{equation}
In the equations above, N$_{uv}$, c$_{s}$ \citep[= 11 km s$^{-1}$; in the ionized gas;][]{bisbas09}, and $\alpha_{B}$ are previously defined (see Section~\ref{subsec:radio}), 
$\mu$= 0.678 \citep[in the ionized gas;][]{bisbas09}, m$_{H}$ is the hydrogen atom mass, $\dot{M}_{w}$ is the mass-loss rate, 
V$_{w}$ is the wind velocity of the ionizing source,  L$_{bol}$ is the bolometric luminosity of the source, and 
D$_{s}$ is the projected distance from the location of the B0.5V star where the pressure components are estimated. 
The pressure components driven by a massive star are evaluated at D$_{s}$ = 3.5 pc 
(i.e. the separation between the positions of the NVSS peak and clump1). 

Adopting, $L_{bol}$ = 19952 L$_{\odot}$ \citep[for a B0.5V star;][]{panagia73}, $\dot{M}_{w}$ = 2.5 $\times$ 10$^{-9}$ M$_{\odot}$ yr$^{-1}$ 
\citep[for a B0.5V star;][]{oskinova11}, 
V$_{w}$ = 1000 km s$^{-1}$ \citep[for a B0.5V star;][]{oskinova11} in the above equations, 
we compute 
P$_{HII}$ $\approx$ 2.0 $\times$ 10$^{-11}$ dynes\, cm$^{-2}$, 
$P_{rad}$ $\approx$ 1.7 $\times$ 10$^{-12}$ dynes\, cm$^{-2}$, and
P$_{wind}$ $\approx$ 1.1 $\times$ 10$^{-14}$  dynes\, cm$^{-2}$. 
The comparison of these pressure components indicates that the pressure 
of the H\,{\sc ii} region is relatively higher than the radiation pressure and the stellar wind pressure. 
We also obtain the total pressure (P$_{total}$ = P$_{HII}$ + $P_{rad}$ + P$_{wind}$) driven by a massive star to be 
$\sim$2.2 $\times$ 10$^{-11}$ dynes\, cm$^{-2}$. $P_{total}$ is comparable to the pressure associated with a typical cool molecular 
cloud ($P_{MC}$$\sim$10$^{-11}$--10$^{-12}$ dynes cm$^{-2}$ for a temperature $\sim$20 K 
and particle density $\sim$10$^{3}$--10$^{4}$ cm$^{-3}$) \citep[see Table 7.3 of][]{dyson80}, suggesting that the clump1 is not 
destroyed by the impact of the ionized gas. 
\section{Discussion}
\label{sec:disc}
In recent years, {\it Spitzer} and {\it Herschel} data have observed MIR shells or bubbles, infrared filaments, 
and young star clusters together in many massive star-forming regions, indicating the onset of numerous complex physical processes.
The presence of bubbles/shells associated with H\,{\sc ii} regions is often explained by the feedback mechanisms 
(such as ionizing radiation, stellar winds, and radiation pressure) of massive stars \citep[e.g.,][]{zin07,deharveng10,tan14,dewangan15}. 
The multi-band images have revealed an almost sphere-like shell morphology as the most prominent structure in the S237 region. 
The shell-like HISA feature surrounding the ionized emission is also seen in the 21 cm H\,{\sc i} line data. 
The velocity structure of molecular gas has indicated the presence of an expanding H\,{\sc ii} region associated with the S237 region (see Section~\ref{sec:coem}). 
Additionally, in Section~\ref{sec:feedb}, based on the pressure calculations (P$_{HII}$, $P_{rad}$, and P$_{wind}$), we find that the photoionized gas 
linked with the S237 H\,{\sc ii} region can be considered as the major contributor (against stellar winds and radiation pressure) for the feedback process in the S237 region. Hence, the presence of a bell-shaped cavity-like morphology could be explained by the impact of ionizing photons (see Section~\ref{subsec:h2out}).  

In Section~\ref{subsec:surfden}, we detect only two clusters of YSOs in the S237 region. 
Thirteen {\it Herschel} clumps are identified in the S237 region and star formation activities are exclusively found toward the 
{\it Herschel} clump1 and clump2 (see Section~\ref{subsec:surfden}). 
Interestingly, the clump1 and clump2 contain the bell-shaped cavity-like structure hosting the peaks of 1.4 GHz emission as well as diffuse H$\alpha$ emission
and filamentary features without any radio continuum detection, respectively. 
\citet{lim15} also found two groups of PMS stars inferred using their surface density analysis 
and estimated a median age of PMS members to be 2.0 Myr. They also reported the maximum age difference between the
stars in the these two groups to be about 0.7 Myr, on average.
Considering the presence of the expanding H\,{\sc ii} region, the spatial locations of the YSO clusters may indicate the triggered star formation scenario 
in the S237 region. One can obtain more details about different processes of triggered star formation in the review article by \citet{elmegreen98}.
\citet{evans09} reported an average age of the Class~I and Class~II YSOs to be $\sim$0.44 Myr and $\sim$1--3 Myr, respectively. 
Comparing these typical ages of YSOs with the dynamical age of the S237 H\,{\sc ii} region (i.e. $\sim$0.2--0.8 Myr; see Section~\ref{subsec:radio}), 
it seems that the S237 H\,{\sc ii} region is too young for initiating the formation of a new generation of stars. 
It is also supported by the results obtained by \citet{lim15} (see above).
Hence, the young clusters are unlikely to have been the product of triggered formation. 
The $^{13}$CO emission is not detected toward the clump2 and subreg1, 
indicating the absence of dense gas toward these subregions (see Figure~\ref{fig8a}b). 
Note that the hot gas emission is traced toward the subreg1 using the ROSAT X-ray image and is found away (about 2$\arcmin$) 
from the 1.4 GHz peak emission. The surface density contours are also seen toward the subreg1. 
We suspect that the diffuse X-ray emission could be originated from young stars present in the cluster (see subreg1 in Figure~\ref{fig4a}a).
Due to coarse resolution of the ROSAT X-ray image, we cannot identify the X-ray-emitting young stars. 
In general, \citet{elmegreen11} mentioned that open cluster complexes 
could be the remnants of star formation in giant clouds formed by gravitational instabilities in the Milky Way gas layer. 

It is also worth mentioning that the {\it Herschel} clump1 is the most massive with about 260 M$_{\odot}$ and 
contains the filamentary features and has a noticeable velocity gradient in both the $^{12}$CO and $^{13}$CO emissions.
In Figure~\ref{fig11a}, there is a convincing evidence for a physical association of a cluster of YSOs and a massive clump with the filamentary features, 
indicating the role of filaments in the star formation process. 
Recently, \citet{nakamura14} studied the filamentary ridges in 
the Serpens South Infrared dark cloud using the molecular line observations and 
argued that the filamentary ridges were appeared to converge toward the protocluster clump. Furthermore, they suggested 
the collisions of the filamentary ridges may have triggered cluster formation.
\citet{schneider12} also carried out {\it Herschel} data analysis toward the Rosette Molecular Cloud and suggested that 
the infrared clusters were preferentially seen at the junction of filaments or filament mergers.
They also reported that their outcomes are in agreement with the results obtained in the simulations of \citet{dale11}.  
In the present work, due to coarse beam sizes of the molecular line data, we cannot directly probe the converging of filaments toward the protocluster clump.
Here, a protocluster clump is referred to a massive clump associated with a cluster of YSOs without any radio continuum emission.
However, our results indicate the presence of a cluster of YSOs and a massive clump at the intersection of filamentary features, 
similar to those found in Serpens South region. Therefore, it seems that the collisions of these features may have influenced the cluster formation. 
Based on these indicative outcomes, further detailed investigation of this region is encouraged using high-resolution molecular line observations.

In general, the information of the relative orientation of the mean field direction and the filamentary features offers to infer 
the role of magnetic fields in the formation and evolution of the filamentary features. 
Using previous optical polarimetric observations \citep[from][]{pandey13}, we find an average value of the equatorial position angle of three stars 
to be 161$\degr$.5, which are located near the filamentary features (see the positions of these stars in Figure~\ref{fig11a}b), while the equatorial position angle of the filamentary 
features at their intersection zone is computed to be about 95$\degr$. 
The polarization vectors of background stars indicate the magnetic field direction in the plane of the sky parallel to the direction of
polarization \citep{davis51}. Hence, these filamentary features seem to be nearly perpendicular to the plane-of-the-sky projection of the magnetic 
field linked with the molecular condensation, conds1 in the S237 region, indicating that the magnetic field is likely to have influenced the formation of the filamentary features. 
\citet{sugitani11} studied NIR imaging polarimetry toward the Serpens South cloud and found that the magnetic field is nearly perpendicular 
to the main filament. In particular, the filamentary features in the S237 region appear to be originated by a similar process as observed in 
the Serpens South cloud. However, high-resolution polarimetric observations at longer wavelengths will be helpful to further explore the role of magnetic field in the formation of the filamentary features in the S237 region.
\section{Summary and Conclusions}
\label{sec:conc}
In order to investigate star formation processes in the S237 region, we have utilized multi-wavelength data 
covering from radio, NIR, optical H$\alpha$ to X-ray wavelengths. Our analysis has been focused on the molecular gas kinematics, ionized emission, hot gas, cold dust emission, 
and embedded young populations. Our main findings are as follows:\\\\ 
$\bullet$ The S237 region has a broken or incomplete ring or shell-like appearance at wavelengths longer than 2 $\mu$m and 
contains a prominent bell-shaped cavity-like morphology at the center, where the peak of the radio-continuum emission is observed. 
The elongated filamentary features are also seen at the edge of the shell-like structure, where the radio-continuum emission is absent. \\
$\bullet$ The distribution of ionized emission traced in the NVSS 1.4 GHz continuum map is almost spherical 
and the S237 H\,{\sc ii} region is powered by a radio spectral type of B0.5V star. The dynamical age of the S237 H\,{\sc ii} region is 
estimated to be $\sim$0.2(0.8) Myr for 10$^{3}$(10$^{4}$) cm$^{-3}$ ambient density.\\
$\bullet$ The molecular cloud associated with the S237 region (i.e. S237 molecular cloud) is well 
traced in a velocity range of $-$7 to $-$2 km s$^{-1}$. In the integrated $^{12}$CO map, at least four 
molecular condensations (conds1-4) are identified.\\
$\bullet$ In the integrated $^{13}$CO map, the molecular gas is seen only toward the condensation, conds1.\\
$\bullet$ Using 0.5--2 keV X-ray image, the hot gas emission is traced at the center of the shell-like morphology and could be due 
to young stars present in the cluster. \\
$\bullet$ The position-velocity analysis of $^{12}$CO emission depicts an inverted C-like structure, 
revealing the signature of an expanding H\,{\sc ii} region with a velocity of $\sim$1.65 km s$^{-1}$.\\ 
$\bullet$ The pressure calculations (P$_{HII}$, $P_{rad}$, and P$_{wind}$) indicate that the photoionized 
gas associated with the S237 H\,{\sc ii} region could be responsible for the origin of the bell-shaped structure seen in the S237 region. \\ 
$\bullet$ Thirteen {\it Herschel} clumps have been traced in the {\it Herschel} column density map. 
The majority of molecular gas is distributed toward a massive {\it Herschel} clump1 (M$_{clump}$ $\sim$260 M$_{\odot}$), 
which contains the filamentary features. 
The position-velocity analysis of $^{12}$CO and $^{13}$CO emissions traces a noticeable velocity gradient along this {\it Herschel} clump1.\\ 
$\bullet$ The analysis of NIR (1--5 $\mu$m) photometry provides a total of 98 YSOs and also traces the clusters of YSOs 
mainly toward the bell-shaped structure and the filamentary features.\\
$\bullet$ Toward the elongated filamentary features, a cluster of YSOs is spatially coincident with a massive {\it Herschel} clump1 embedded 
within the molecular condensation, conds1.\\

Taking into account the lower dynamical age of the H\,{\sc ii} region (i.e. 0.2-0.8 Myr), the clusters of YSOs are unlikely to be originated by the expansion of 
the H\,{\sc ii} region. An interesting outcome of this work is the existence of a cluster of YSOs 
and a massive clump at the intersection of filamentary features, indirectly illustrating that the collisions of these features 
may have triggered cluster formation, similar to those seen in the Serpens South star-forming region. 
\acknowledgments
We thank the anonymous reviewer for a critical reading of the manuscript and several useful comments and 
suggestions, which greatly improved the scientific contents of the paper. 
The research work at Physical Research Laboratory is funded by the Department of Space, Government of India. 
This work is based on data obtained as part of the UKIRT Infrared Deep Sky Survey. This publication 
made use of data products from the Two Micron All Sky Survey (a joint project of the University of Massachusetts and 
the Infrared Processing and Analysis Center / California Institute of Technology, funded by NASA and NSF), archival 
data obtained with the {\it Spitzer} Space Telescope (operated by the Jet Propulsion Laboratory, California Institute 
of Technology under a contract with NASA). 
The Canadian Galactic Plane Survey (CGPS) is a Canadian project with international partners. 
The Dominion Radio Astrophysical Observatory is operated as a national facility by the 
National Research Council of Canada. The Five College Radio Astronomy Observatory 
CO Survey of the Outer Galaxy was supported by NSF grant AST 94-20159. The CGPS is 
supported by a grant from the Natural Sciences and Engineering Research Council of Canada. 
This paper makes use of data obtained as part of the INT Photometric H$\alpha$ Survey of the Northern Galactic 
Plane (IPHAS, www.iphas.org) carried out at the Isaac Newton Telescope (INT). The INT is operated on the 
island of La Palma by the Isaac Newton Group in the Spanish Observatorio del Roque de los Muchachos of 
the Instituto de Astrofisica de Canarias. All IPHAS data are processed by the Cambridge Astronomical Survey 
Unit, at the Institute of Astronomy in Cambridge.
IZ is supported by the Russian Foundation for Basic Research (RFBR). 
AL acknowledges the CONACYT(M\'{e}xico) grant CB-2012-01-1828-41. 
We have made use of the ROSAT Data Archive of the Max-Planck-Institut f\"{u}r extraterrestrische Physik (MPE) at Garching, Germany.
%
\begin{figure*}
\epsscale{0.56}
\plotone{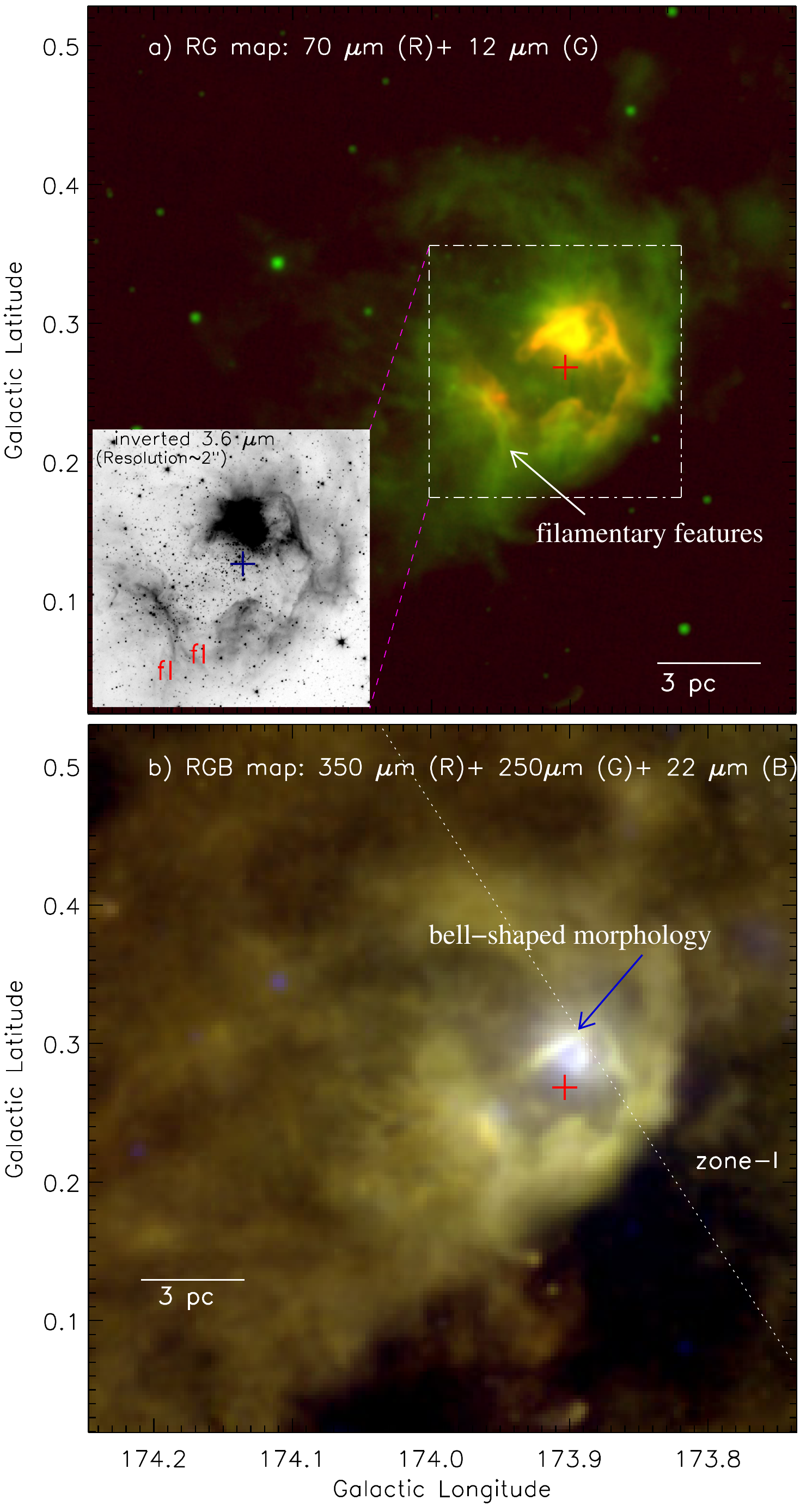}
\caption{\scriptsize Distribution of the MIR and sub-mm emissions toward the S237 region 
(size of the selected region $\sim$30$\farcm6$ $\times$ 30$\farcm6$ ($\sim$20.5 pc $\times$ 20.5 pc at a distance of 2.3 kpc); 
centered at $l$ = 173$\degr$.993; $b$ = 0$\degr$.273). 
a) Two color composite image (in logarithmic gray scale), which clearly illustrates the morphology of the region. 
The inset on the bottom left shows the central region in zoomed-in view, using the {\it Spitzer} 3.6 $\mu$m image 
(inverted gray scale; see a dotted-dashed white box in figure). Two filamentary features are highlighted by labels ``fl".
b) Composite trichromatic image of the S237 region with the {\it Herschel} 350 $\mu$m, 250 $\mu$m, and {\it WISE} 22 $\mu$m bands in red, green, and blue, respectively. The UKIDSS GPS data are available only for the zone-I area as highlighted by a dotted-line. 
In both the panels, the position of IRAS 05281+3412 is marked by a plus (``+") symbol. 
In both the panels, the scale bar corresponding to 3 pc (at a distance of 2.3 kpc) is shown in bottom-left. 
Figures reveal a shell-like morphology of the region, filamentary features, and a bell-shaped cavity-like structure.}
\label{fig1a}
\end{figure*}
\begin{figure*}
\epsscale{0.59}
\plotone{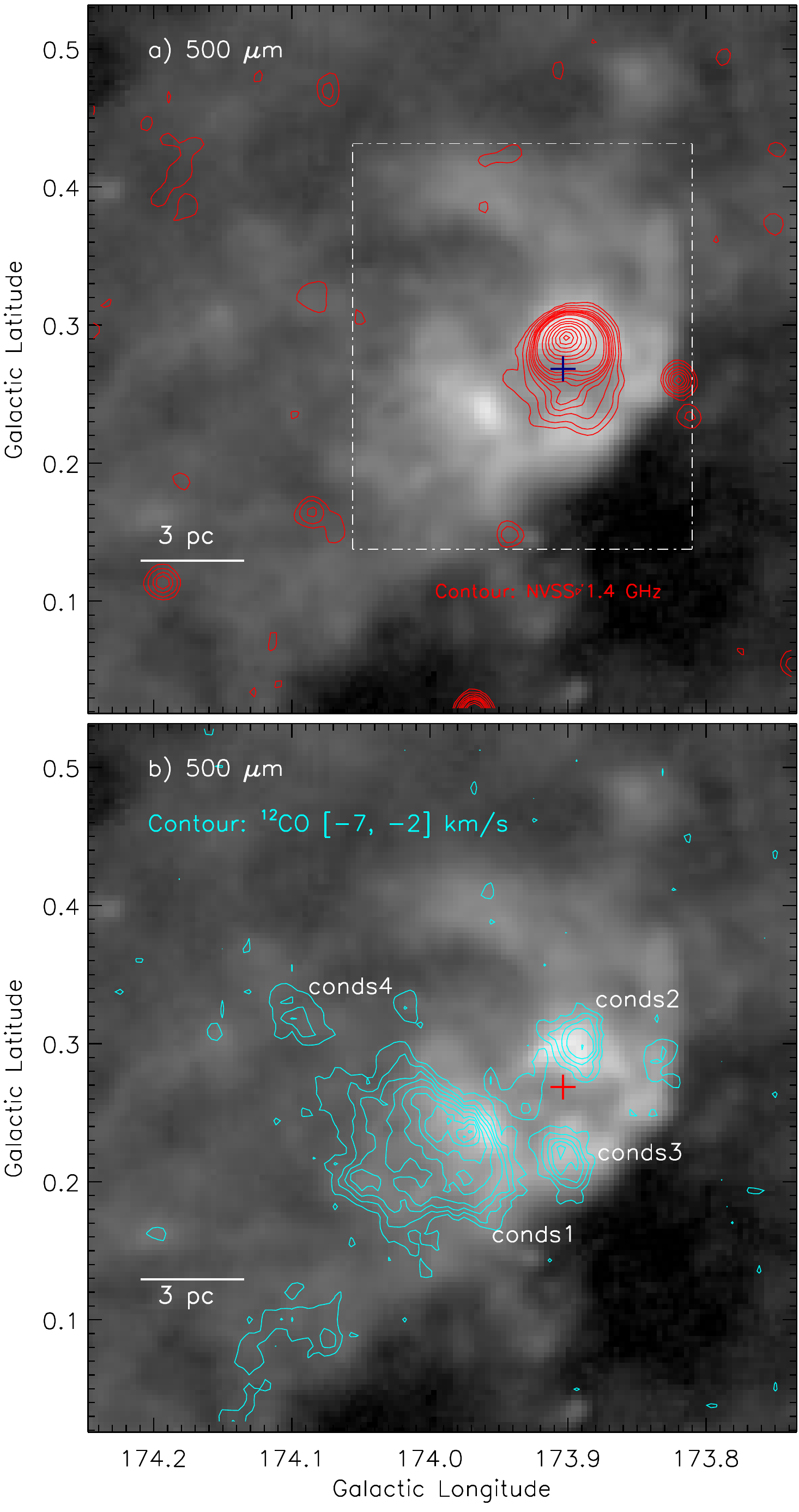}
\caption{\scriptsize Distribution of the sub-mm, ionized, and molecular emissions toward the S237 region.
 a) The radio continuum contours from the NVSS at 1.4 GHz are overlaid on the {\it Herschel} 500 $\mu$m image. 
 The NVSS contours are superimposed with levels of 0.5, 1, 2, 3, 4, 5, 6, 8, 10, 20, 30, 40, 55, 70, 85, and 95\% of 
the peak value (i.e., 0.1936 Jy/beam). The dotted-dashed white box encompasses the area shown in Figure~\ref{fig3a}a. 
 b) The integrated $^{12}$CO (1--0) emission contours are overplotted on the {\it Herschel} 500 $\mu$m image.
The $^{12}$CO emission contours are superimposed with levels of 1, 2, 3, 4, 55, 70, 80, 90, and 99\% of 
the peak value (i.e., 36.5639 K km s$^{-1}$). 
In each panel, the scale bar at the bottom-left corner corresponds to 3 pc (at a distance of 2.3 kpc). 
In each panel, the marked symbol is similar to those shown in Figure~\ref{fig1a}. 
Figures show the distribution of ionized emission, molecular gas, and dense materials in the region. }
\label{fig2a}
\end{figure*}
\begin{figure*}
\epsscale{0.54}
\plotone{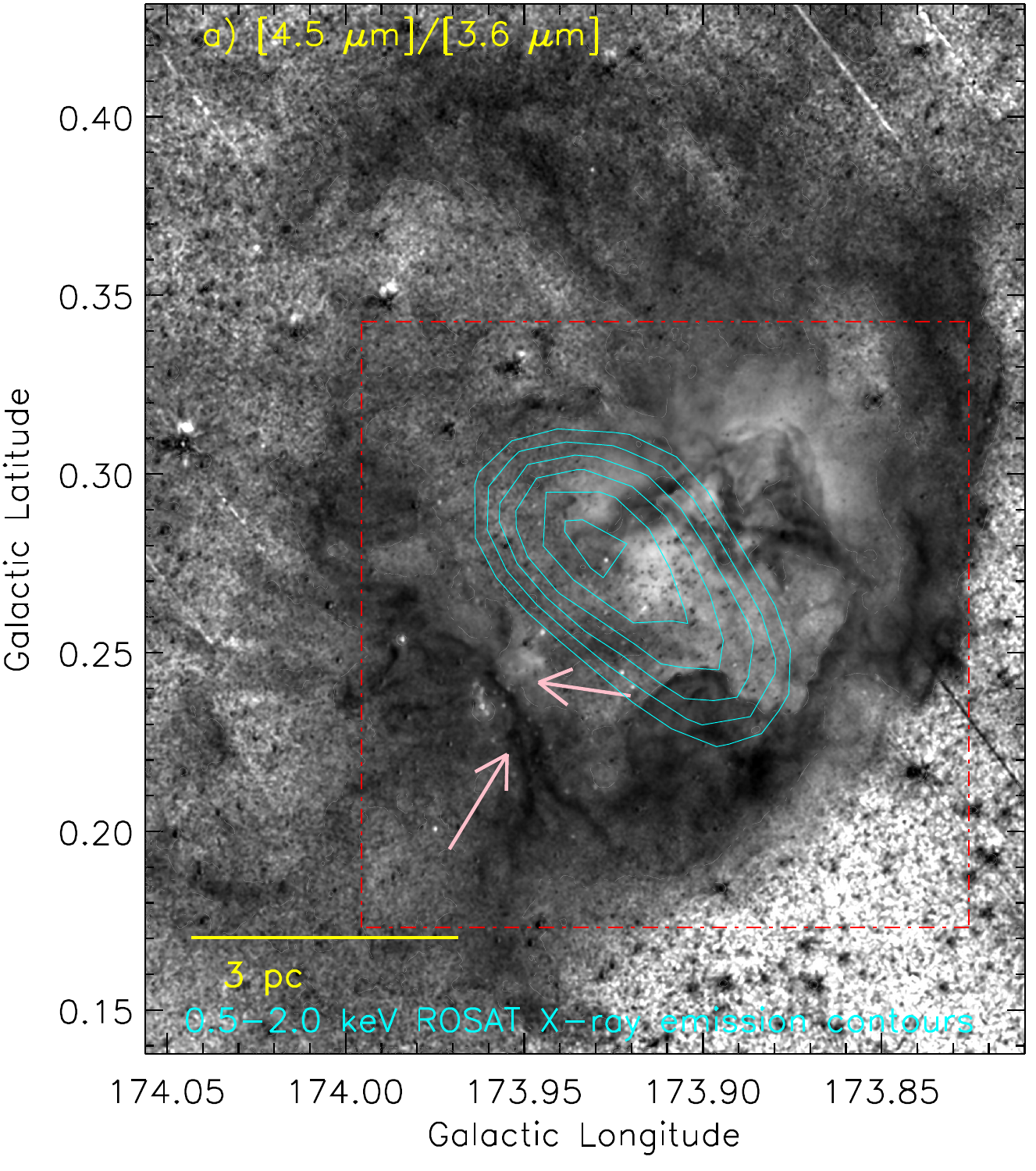}
\epsscale{0.51}
\plotone{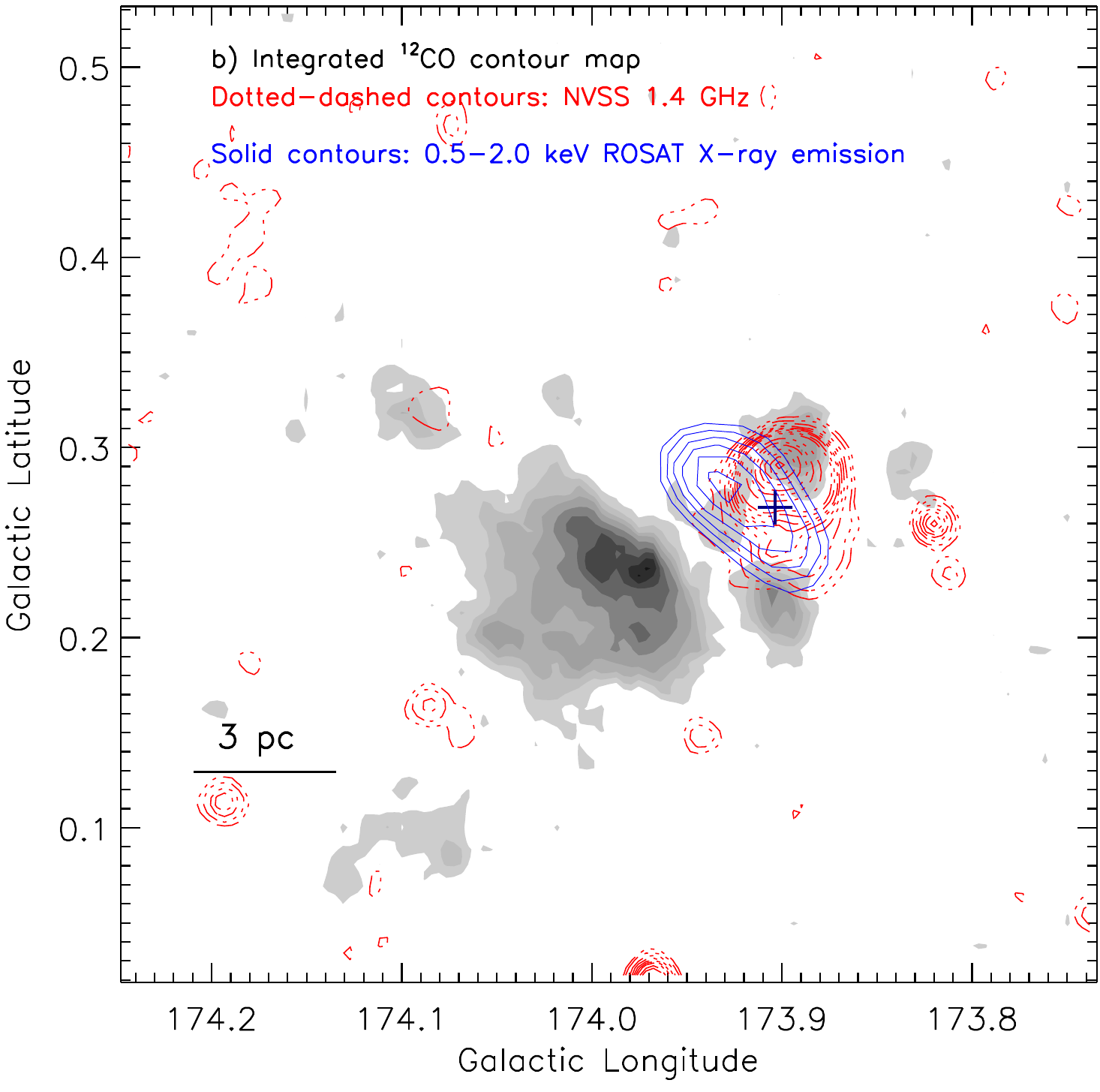}
\caption{\scriptsize a) {\it Spitzer} ratio map of 4.5 $\mu$m/3.6 $\mu$m emission. 
The ratio map is subjected to median filtering with a width of 5 pixels and smoothened by 
4 pixel $\times$ 4 pixel using a boxcar algorithm. The dotted-dashed box (in red) encompasses the 
area shown in Figure~\ref{fig4af}. 
The 4.5 $\mu$m/3.6 $\mu$m emission contour (gray dotted-dashed contour) is also shown on the image with a representative value of 0.75, 
tracing the shell-like morphology. 
b) The contour map of integrated $^{12}$CO emission in the velocity range of $-$7 to $-$2 km s$^{-1}$. 
The contour levels are similar to the one shown in Figure~\ref{fig2a}b. 
The NVSS contours (red dotted-dashed contours) are also superimposed with similar levels to that shown in Figure~\ref{fig2a}a. 
The 0.5--2.0 keV ROSAT X-ray emission contours (solid contours) of S237 are overplotted on both the panels. 
The X-ray image is Gaussian smoothed with a radius of 3 pixels. 
In both the panels, the scale bar at the bottom-left corner corresponds to 3 pc (at a distance of 2.3 kpc).}
\label{fig3a}
\end{figure*}
\begin{figure*}
\epsscale{1}
\plotone{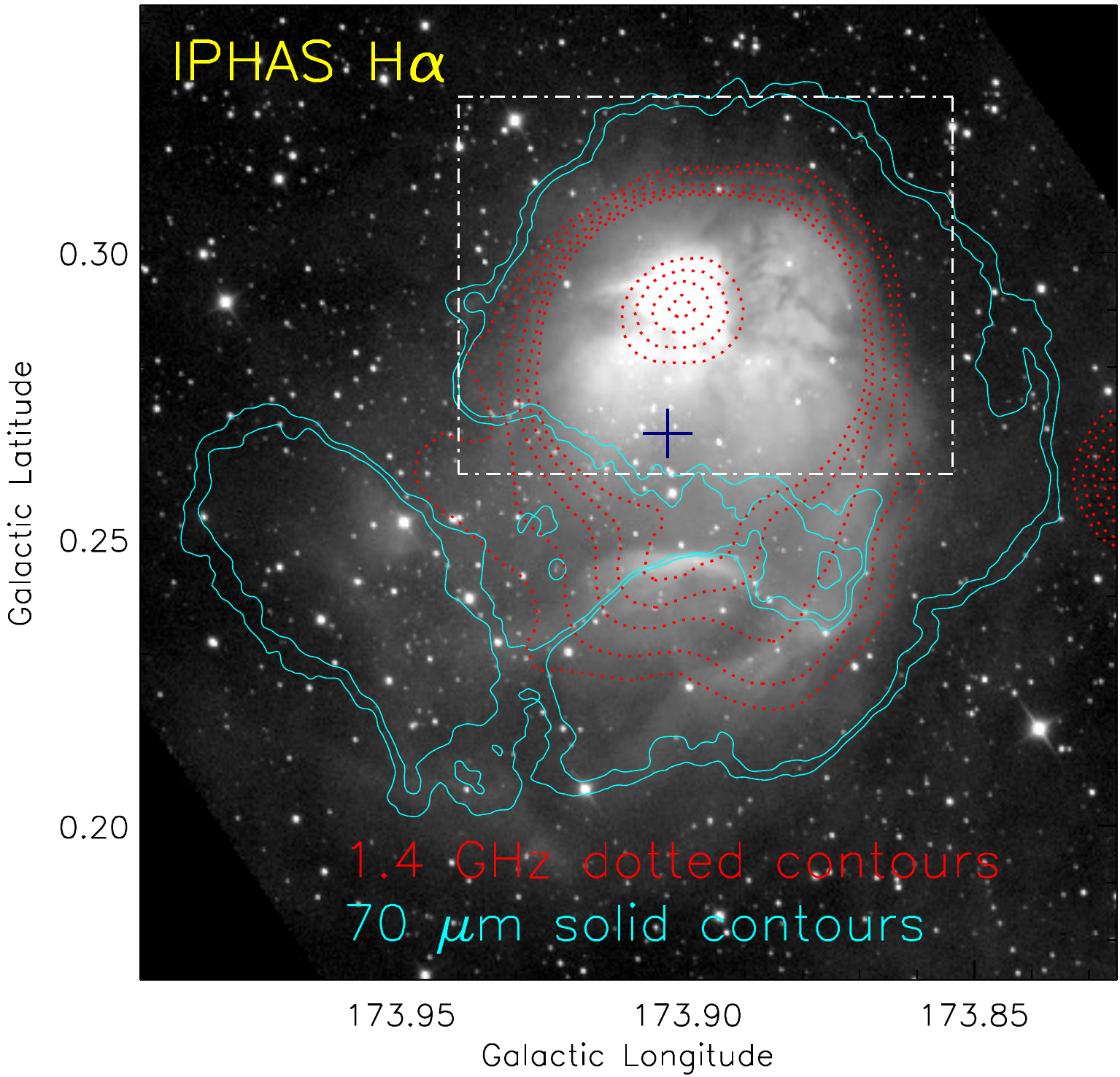}
\caption{\scriptsize IPHAS H$\alpha$ gray-scale image is overlaid with the NVSS 1.4 GHz and {\it Herschel} 70 $\mu$m emission contours. 
The NVSS 1.4 GHz dotted contours (in red) are shown with levels of 0.5, 1, 2, 3, 4, 55, 70, 85, and 95\% of the 
peak value (i.e. 0.1936 Jy/beam). 
The dotted-dashed white box encompasses the area shown in Figures~\ref{fig4a}a and~\ref{fig4a}b.}
\label{fig4af}
\end{figure*}
\begin{figure*}
\epsscale{0.8}
\plotone{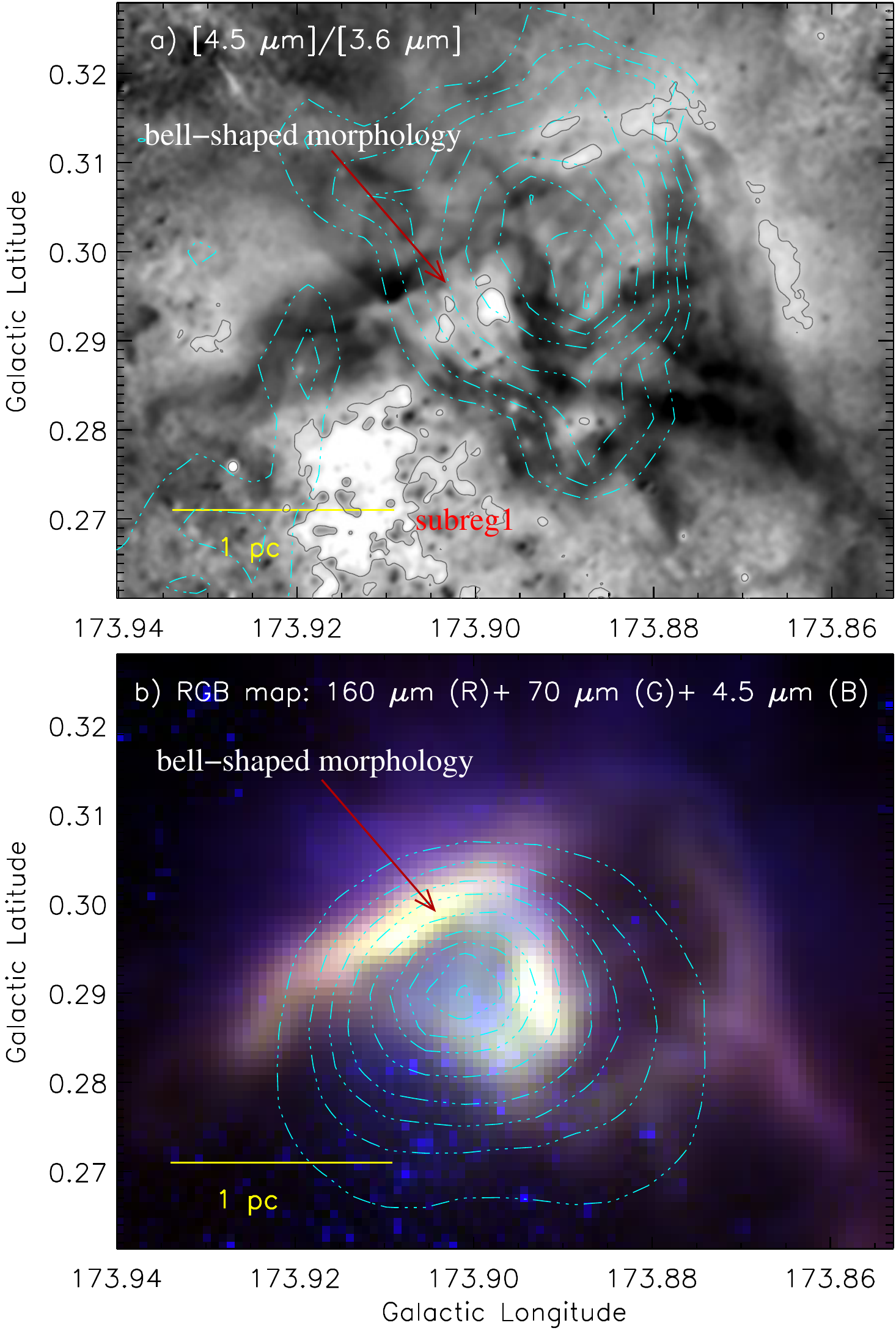}
\caption{\scriptsize a) {\it Spitzer} ratio map of 4.5 $\mu$m/3.6 $\mu$m emission is overlaid with the integrated $^{12}$CO (J = 1-0) contours. 
The $^{12}$CO contours are shown with levels of 20, 30, 40, 55, 70, 80, 90, and 99\% of the peak value (i.e. 20.076 K km s$^{-1}$). 
The 4.5 $\mu$m/3.6 $\mu$m emission contour (in dark gray) is also shown on the image with a representative value of 0.99. 
b) Color-composite map using 160 $\mu$m (red), 70 $\mu$m (green), and 4.5 $\mu$m (blue) images.
The map is overlaid with the NVSS 1.4 GHz contours. 
The NVSS 1.4 GHz contours are shown with levels of 
10, 20, 30, 40, 55, 70, 80, 90, and 99\% of the peak value (i.e. 0.1936 Jy/beam). 
In each panel, the scale bar at the bottom-left corner corresponds to 1 pc (at a distance of 2.3 kpc).}
\label{fig4a}
\end{figure*}
\begin{figure*}
\epsscale{1}
\plotone{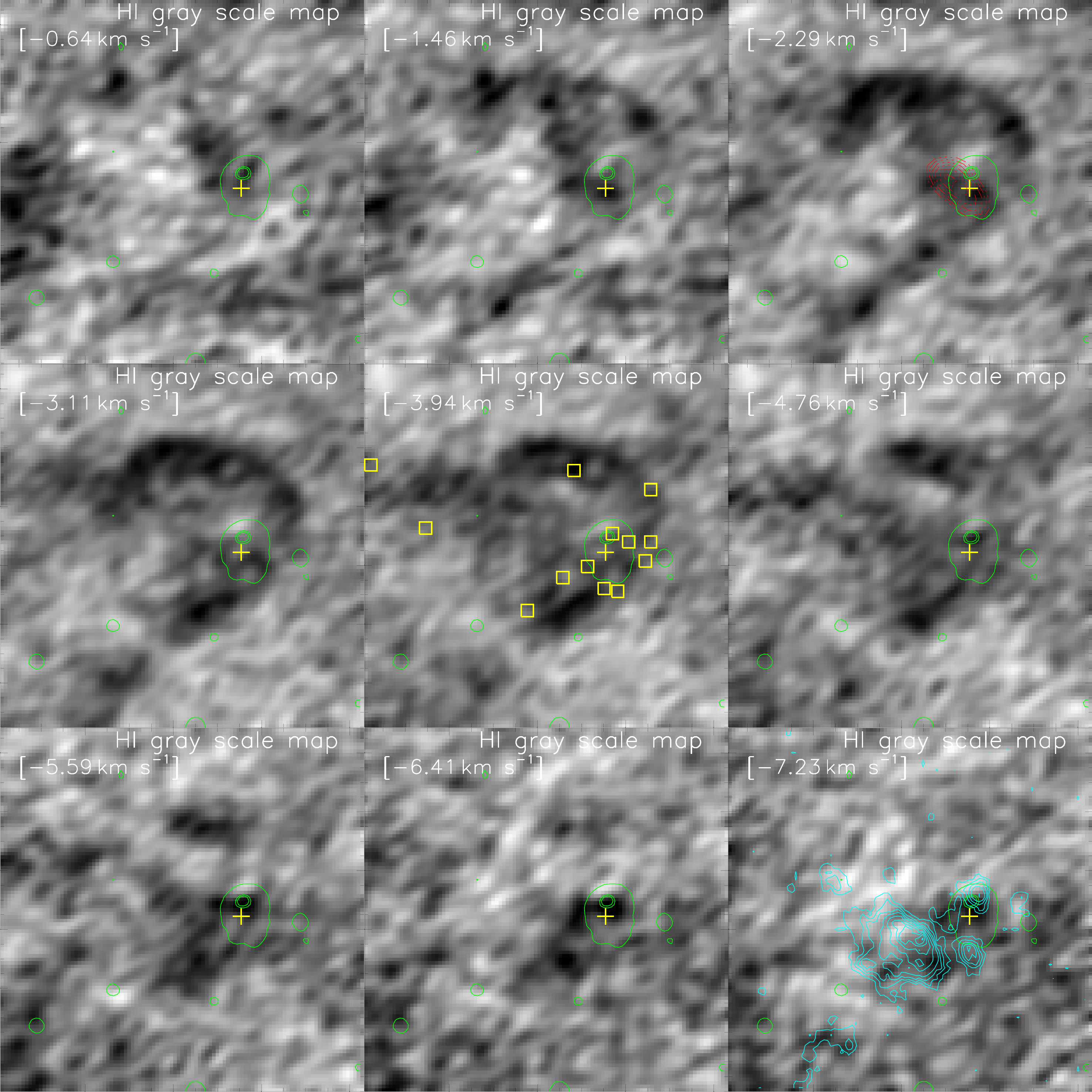}
\caption{\scriptsize The CGPS 21 cm H\,{\sc i} velocity channel maps of the S237 region. 
The velocity value (in km s$^{-1}$) is labeled in each panel. 
In each panel, the NVSS 1.4 GHz contours (in green) are superimposed with levels of 1, 55, and 70\% of 
the peak value (i.e., 0.1936 Jy/beam), tracing the ionized hydrogen emission. In each panel, the position of IRAS 05281+3412 is marked by a plus (``+") symbol. In the third panel (at $-$2.29 km s$^{-1}$), the 0.5--2.0 keV ROSAT X-ray emission contours (red dotted-dashed contours) 
are similar to the one shown in Figure~\ref{fig3a}a. In the fifth panel (at $-$3.94 km s$^{-1}$), the identified {\it Herschel} clumps are marked by square symbols (in yellow; also see Figure~\ref{fig5a}b). To enhance the H\,{\sc i} features, each channel map is smoothed with a boxcar average of 3 pixels width. 
In the last panel, the $^{12}$CO emission contours are similar to the one shown in Figure~\ref{fig2a}b. 
In the maps, the sphere-like shell morphology of the S237 region is revealed as H\,{\sc i} self-absorption features 
\citep[i.e., cold H\,{\sc i} traced in absorption against warmer background H\,{\sc i};][]{kerton05}.}
\label{fig4bb}
\end{figure*}
\begin{figure*}
\epsscale{0.488}
\plotone{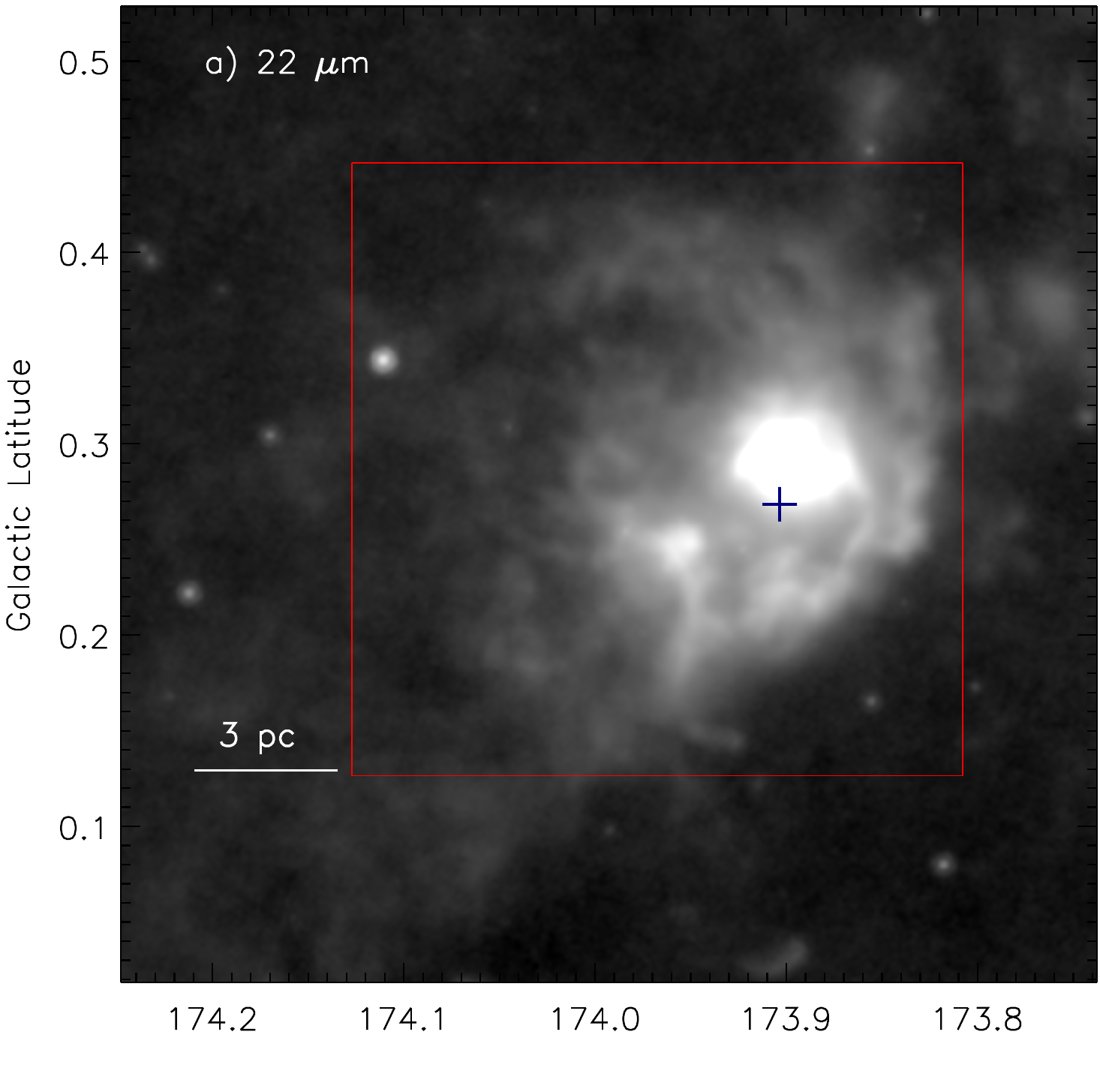}
\epsscale{0.488}
\plotone{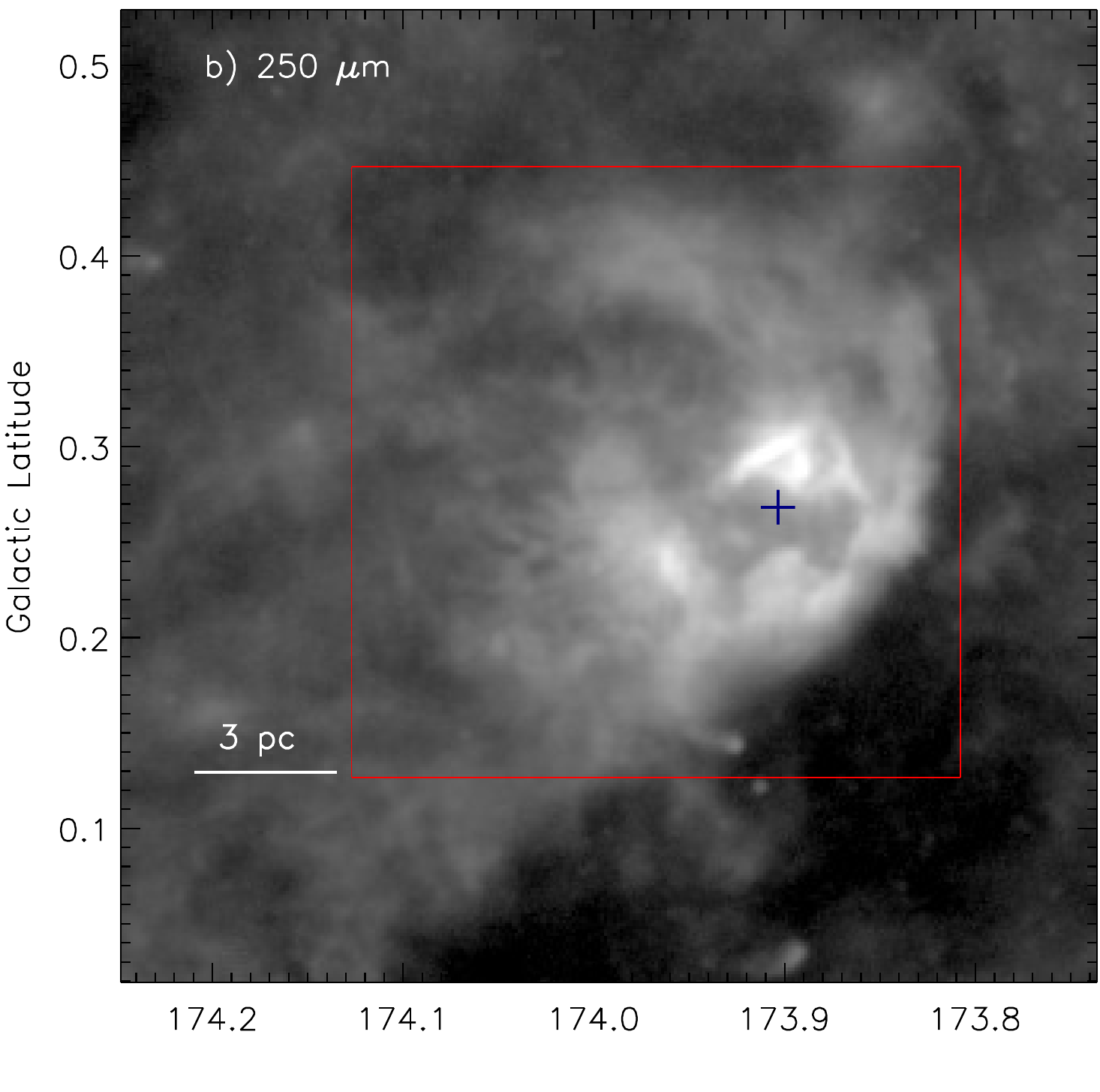}
\epsscale{0.488}
\plotone{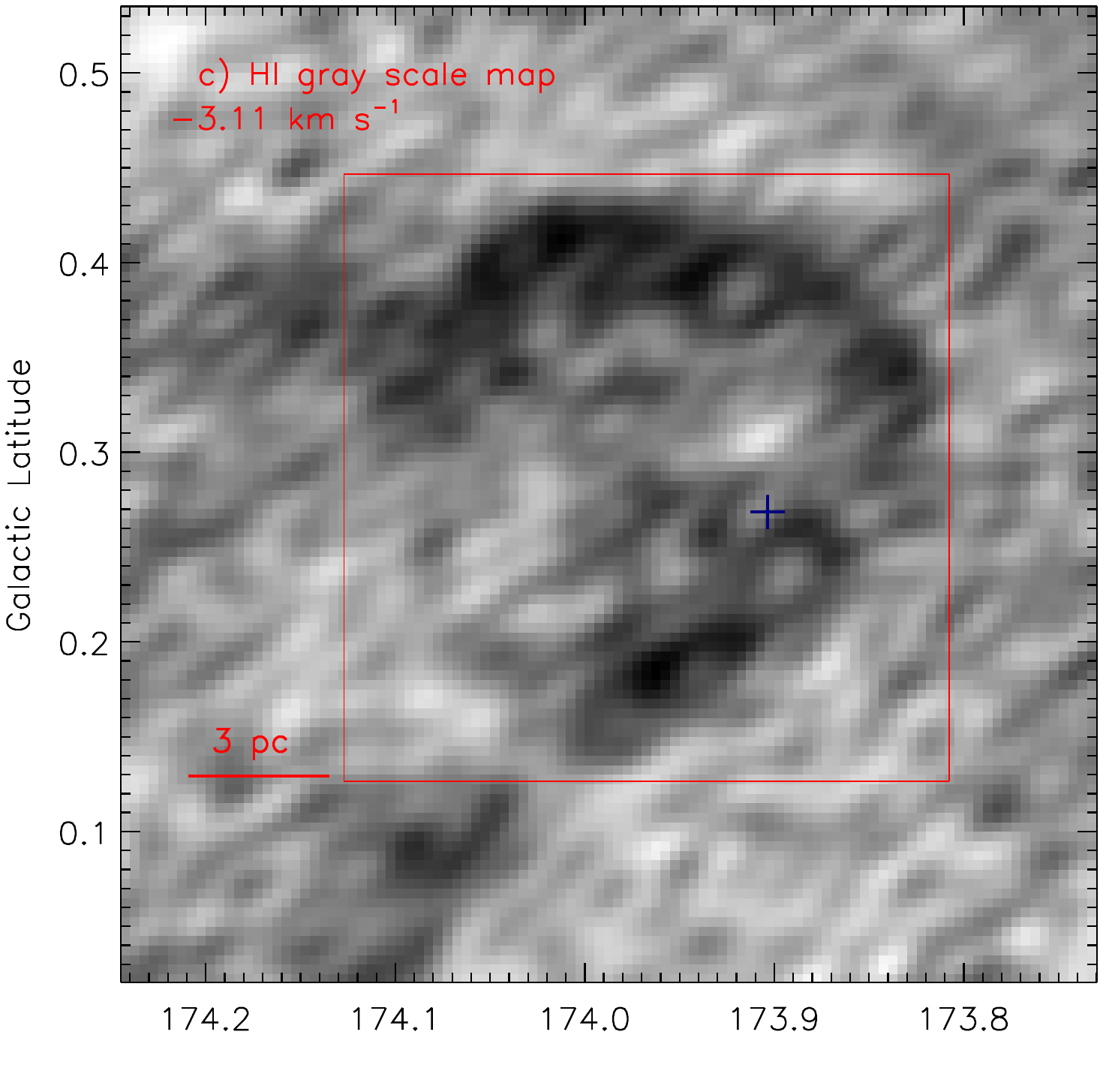}
\caption{\scriptsize Comparison of infrared images and H\,{\sc i} map (at $-$3.11 km s$^{-1}$) of the S237 region. 
In each panel, a sold box (in red) shows an area where the shell-like morphology of the region is seen. 
In each panel, the scale bar at the bottom-left corner corresponds to 3 pc (at a distance of 2.3 kpc).}
\label{fig4uu}
\end{figure*}
\begin{figure*}
\epsscale{0.485}
\plotone{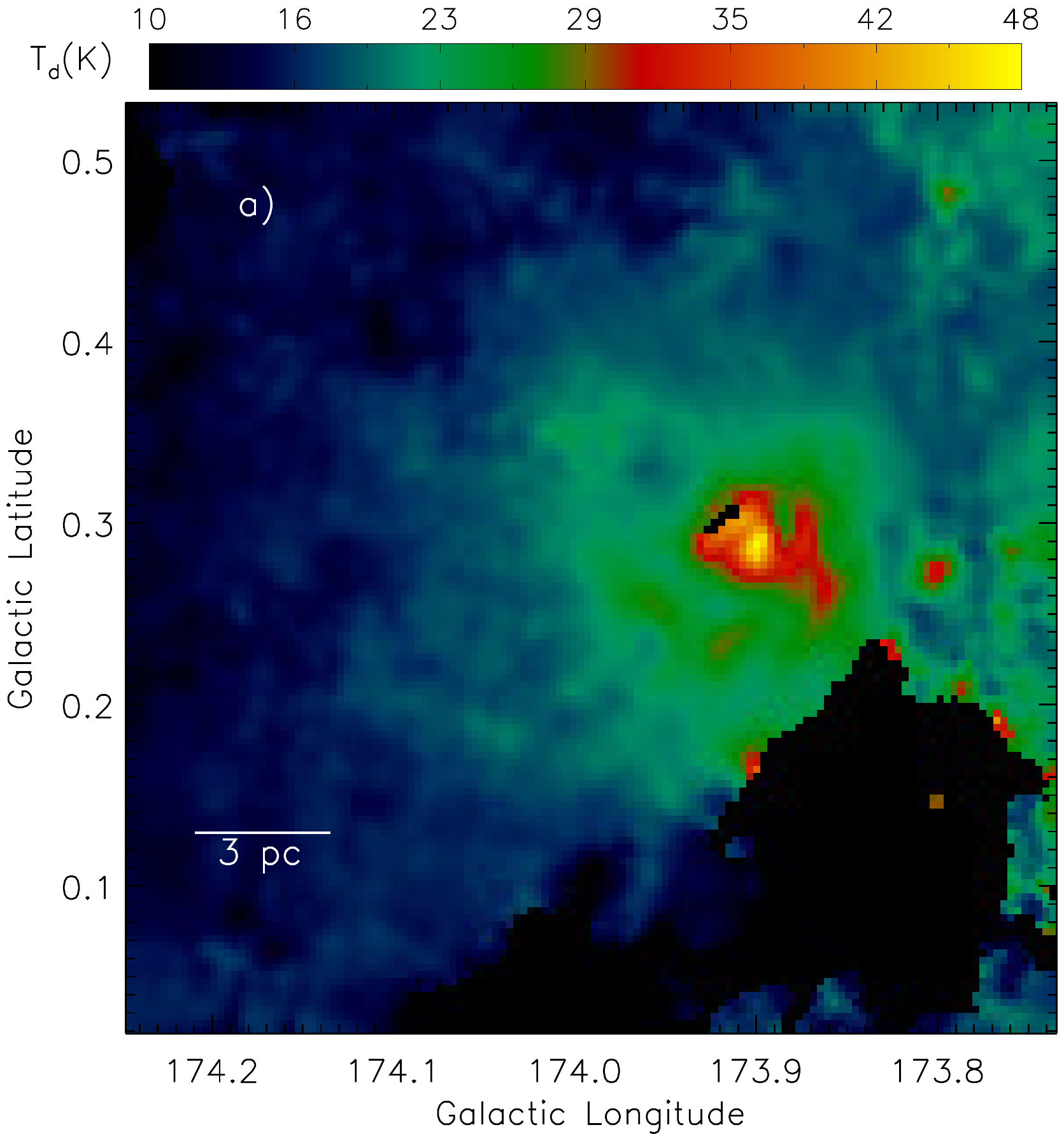}
\epsscale{0.485}
\plotone{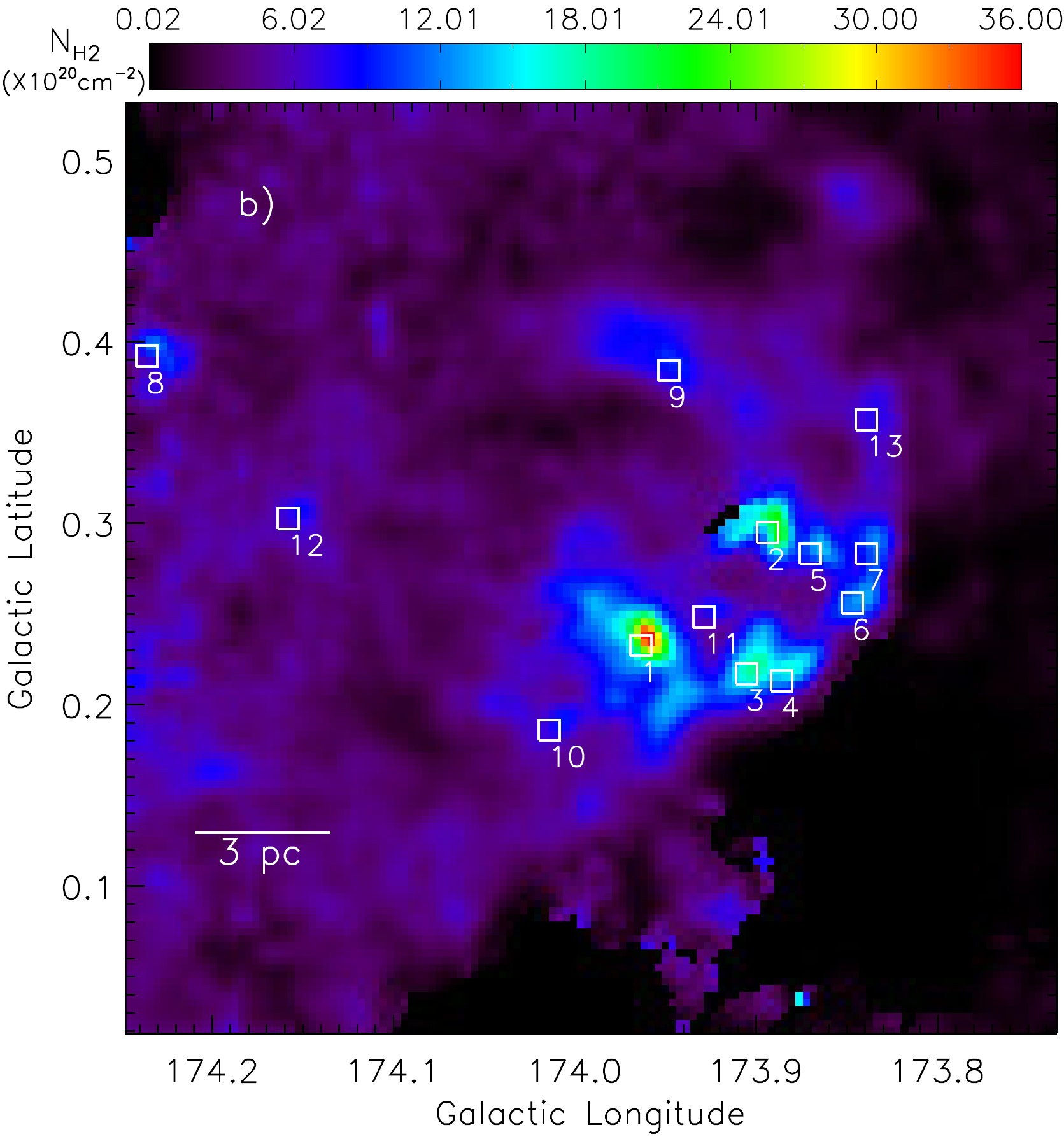}
\epsscale{0.49}
\plotone{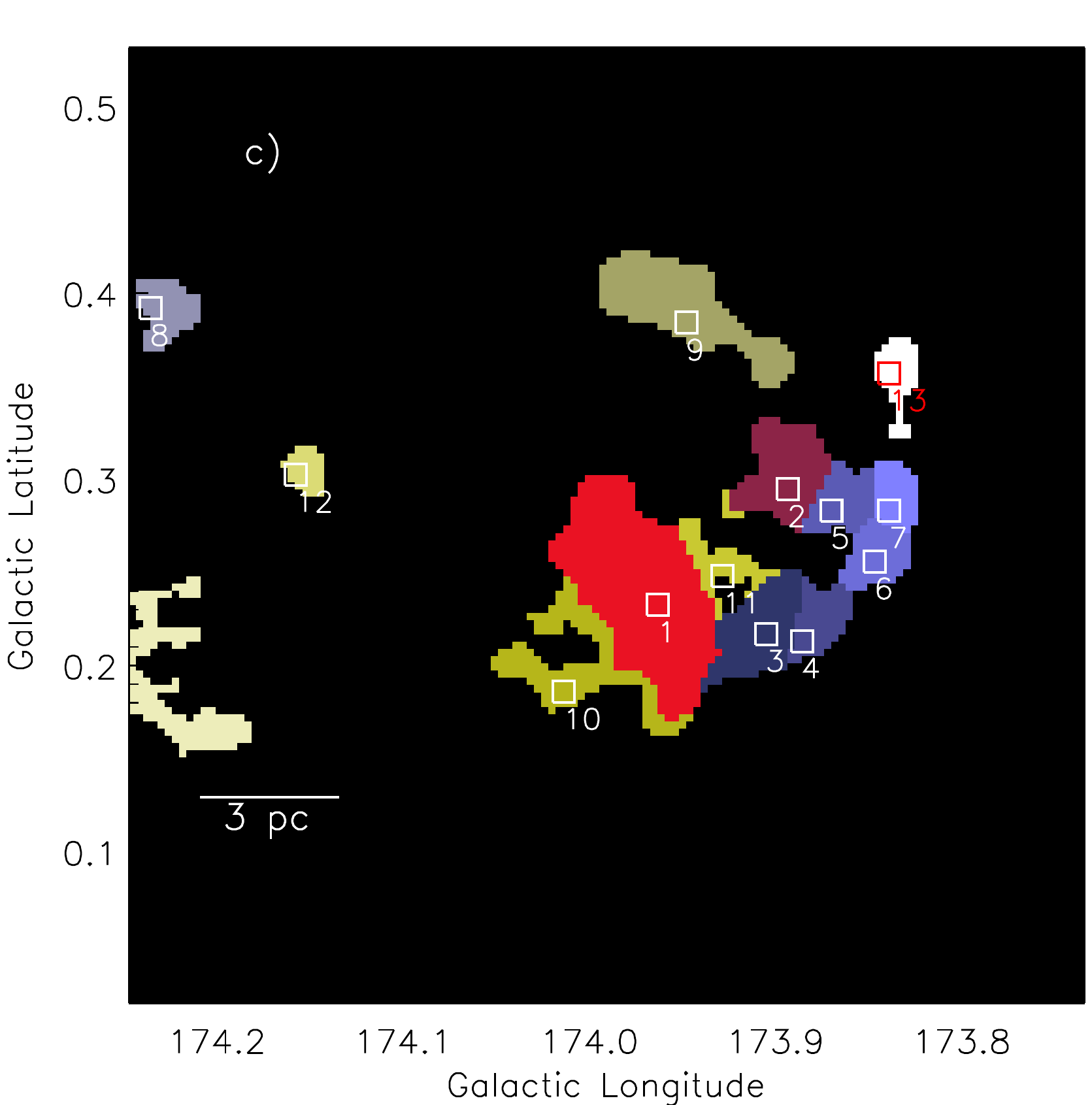}
\caption{\scriptsize {\it Herschel} temperature map (a) and column density ($N(\mathrm H_2)$) map (b) 
of the region (see text for details). The column density map allows to infer the extinction with $A_V=1.07 \times 10^{-21}~N(\mathrm H_2)$ 
and to identify the clumps (see text for details). The identified clumps are marked by square symbols and 
the boundary of each identified clump is shown in Figure~\ref{fig5a}c (also see Table~\ref{tab1}). 
c) The boundary of each identified clump is highlighted along with its corresponding clump 
ID (see Table~\ref{tab1} and also Figure~\ref{fig5a}b). 
In each panel, the scale bar at the bottom-left corner corresponds to 3 pc (at a distance of 2.3 kpc).}
\label{fig5a}
\end{figure*}
\begin{figure*}
\epsscale{1}
\plotone{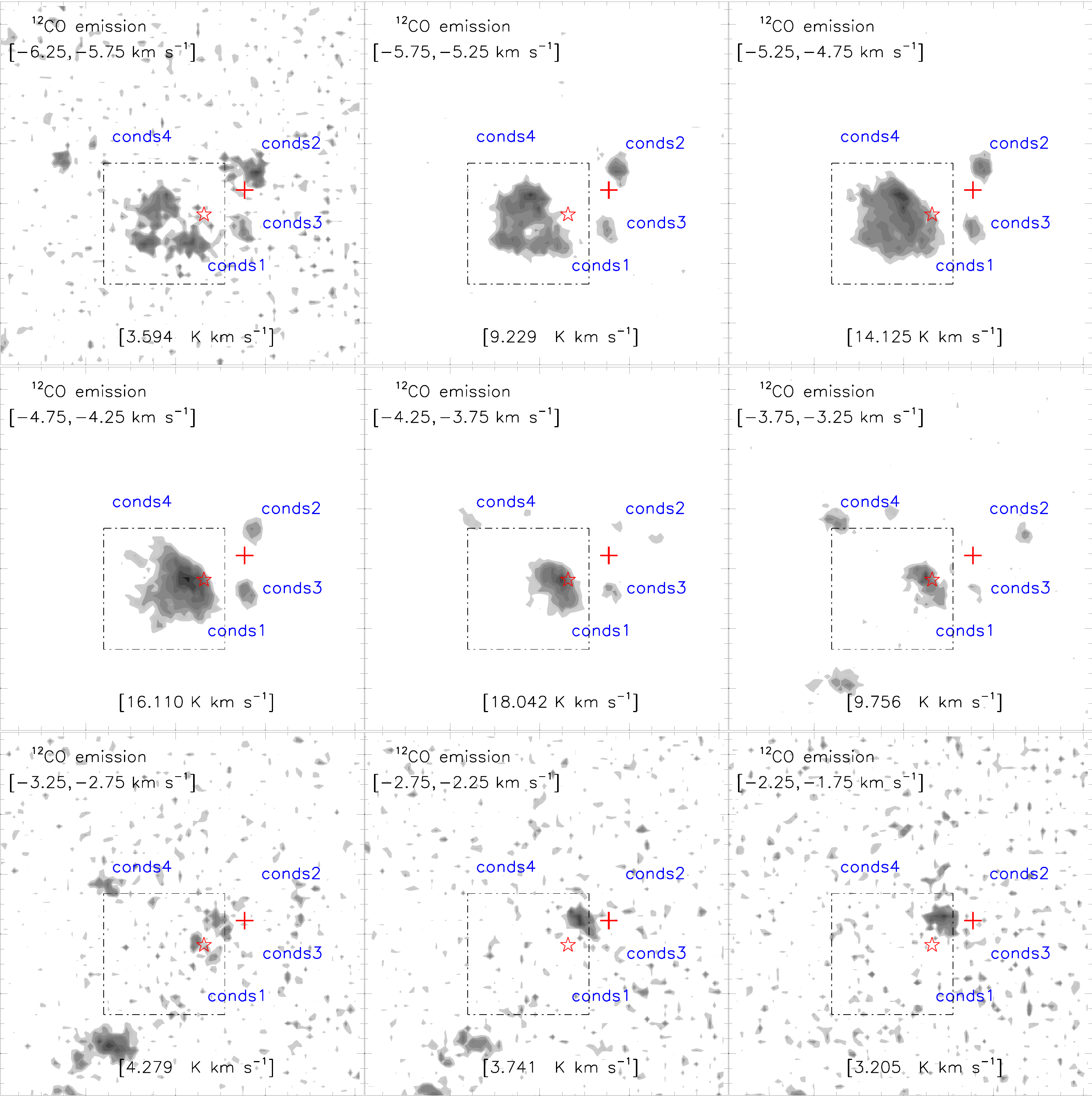}
\caption{\scriptsize Velocity channel contour maps of $^{12}$CO(J =1$-$0) emission. 
The molecular emission is integrated over a velocity interval, which is labeled in each panel (in km s$^{-1}$). 
The contour levels are 18, 25, 30, 40, 55, 70, 80, 90, and 99\% of the peak value (in K km s$^{-1}$), which is also given in each panel.  
In each panel, the positions of IRAS 05281+3412 and {\it Herschel} column density peak are marked 
by a plus (``+") and a star symbols, respectively. 
The dotted-dashed black box encompasses the area shown in Figures~\ref{fig7a} and~\ref{fig11a}. 
In the maps, at least four noticeable condensations (i.e. conds1--4) are observed and the majority of 
molecular gas is found toward the condensation, conds1.}
\label{fig6a}
\end{figure*}
\begin{figure*}
\epsscale{1}
\plotone{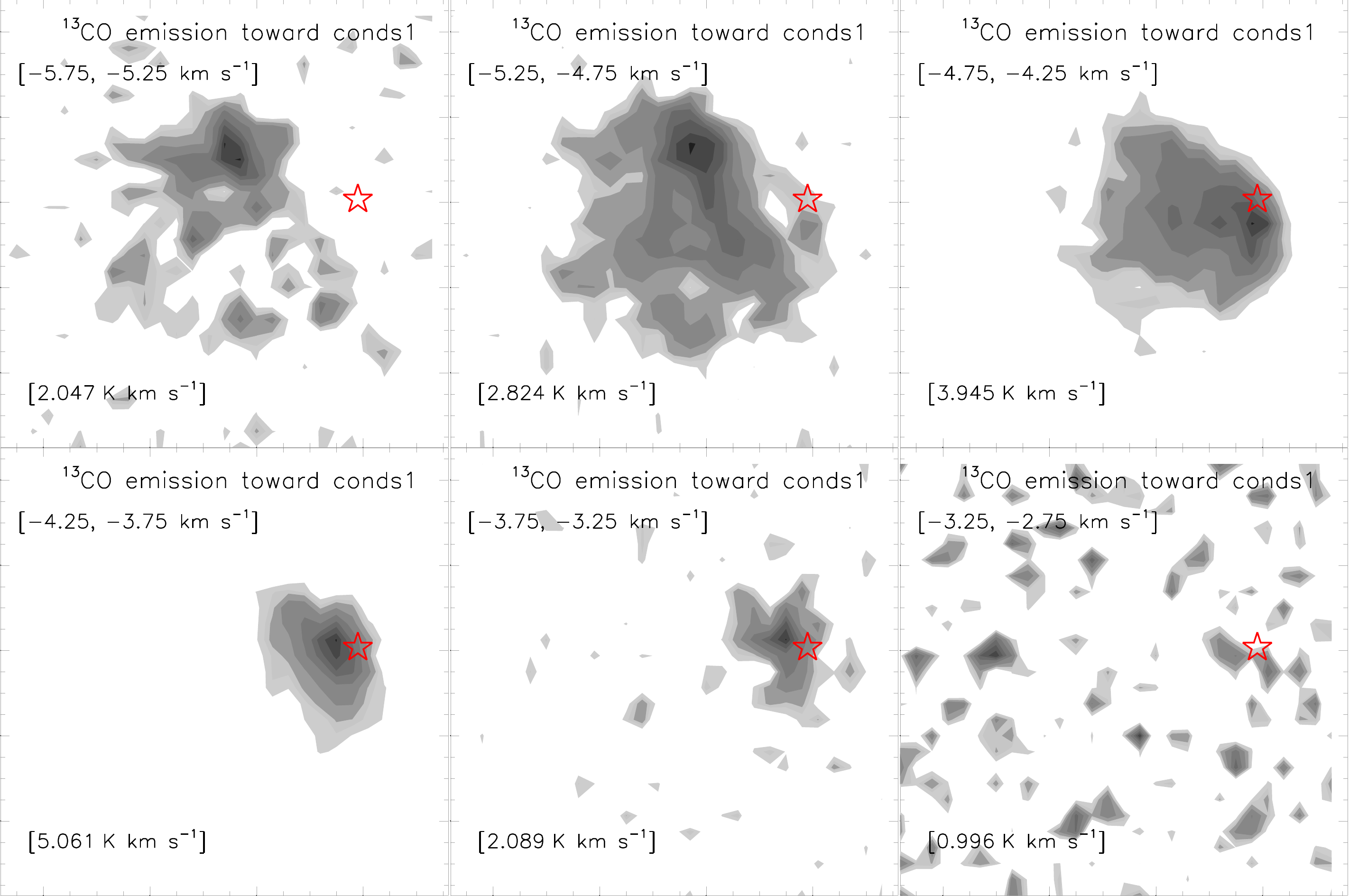}
\caption{\scriptsize The $^{13}$CO(J =1$-$0) velocity channel contour maps toward the molecular condensation, 
conds1 (see dotted-dashed black box in Figure~\ref{fig6a}).
The molecular emission is integrated over a velocity interval, which is labeled in each panel (in km s$^{-1}$). 
The contour levels are 18, 25, 30, 40, 55, 70, 80, 90, and 99\% of the peak value (in K km s$^{-1}$), which is also given in each panel. 
In each panel, the position of the {\it Herschel} column density peak is marked by a star symbol.}
\label{fig7a}
\end{figure*}
\begin{figure*}
\epsscale{0.77}
\plotone{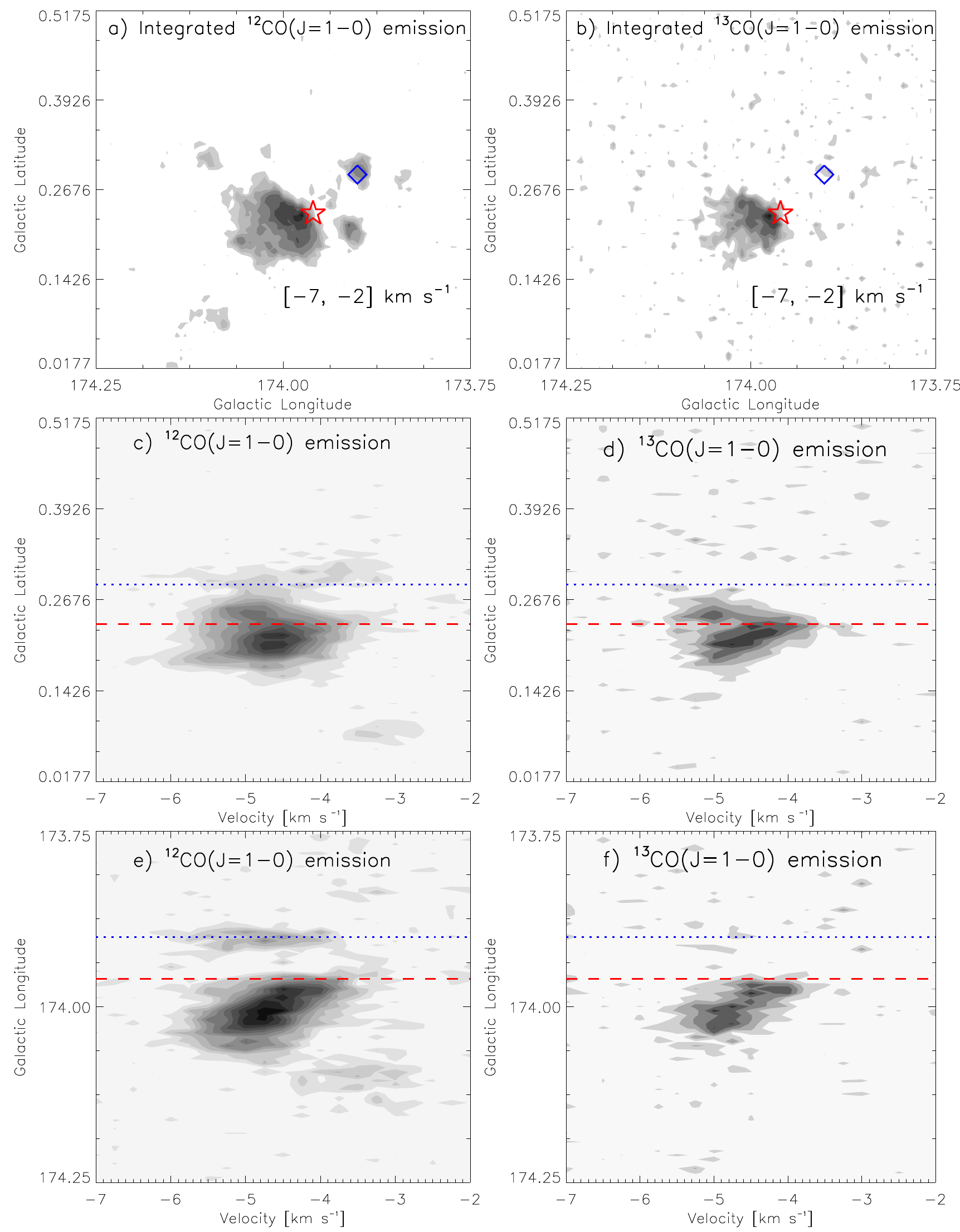}
\caption{\scriptsize The distribution of $^{12}$CO (left) and $^{13}$CO (right) emissions toward the S237 region.
\textbf{top panel:} The contour maps of integrated $^{12}$CO (left; panel ``a") and $^{13}$CO (right; panel ``b") emissions in the velocity range of $-$7 to $-$2 km s$^{-1}$. The $^{12}$CO map is similar to the one shown in Figure~\ref{fig2a}b. 
In top panel, the positions of the NVSS 1.4 GHz peak emission and the {\it Herschel} column density peak are shown by a blue diamond and a red star, respectively. 
\textbf{middle panel:} The Latitude-velocity diagrams of $^{12}$CO (left; panel ``c") and $^{13}$CO (right; panel ``d"). 
In the Latitude-velocity diagrams (panels ``c" and ``d"), the molecular emission is integrated over the longitude from 173$\degr$.75 to 174$\degr$.25.
\textbf{bottom panel:} The Longitude-velocity diagrams of $^{12}$CO (left; panel ``e") and $^{13}$CO (right; panel ``f"). 
In the Longitude-velocity diagrams (panels ``e" and ``f"), the molecular emission is integrated over the latitude from 0$\degr$.0177 to 0$\degr$.5175. 
In the position-velocity diagrams, the dotted blue and dashed red lines show the positions of the NVSS 1.4 GHz peak emission 
and the {\it Herschel} column density peak, respectively. 
In bottom left (panel ``e"), the Longitude-velocity diagram depicts an inverted C-like morphology, indicating the expanding shell with an expanding gas 
velocity of $\sim$1.65 km s$^{-1}$. The position-velocity diagrams of $^{12}$CO and $^{13}$CO also trace a noticeable velocity gradient 
along the {\it Herschel} clump1, which contains the {\it Herschel} column density peak (also see the text).} 
\label{fig8a}
\end{figure*}
\begin{figure*}
\epsscale{1.0}
\plotone{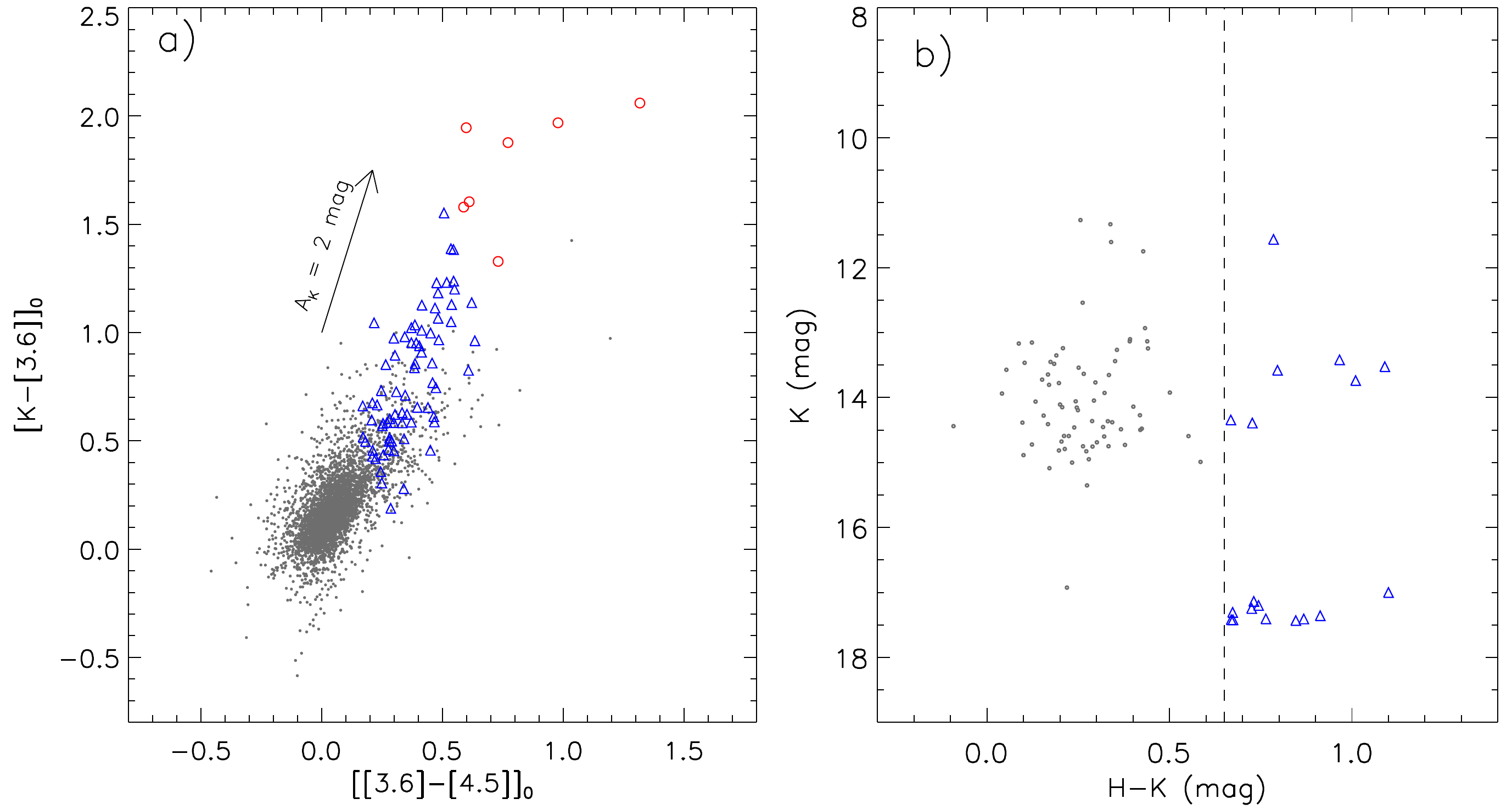}
\caption{\scriptsize Selection of embedded young stellar population within the region probed in this paper (see Figure~\ref{fig1a}). 
a) The dereddened [K$-$[3.6]]$_{0}$ $vs$ [[3.6]$-$[4.5]]$_{0}$ color-color diagram using the H, K, 3.6 $\mu$m, and 4.5 $\mu$m data. 
The extinction vector is drawn using the average extinction laws from \citet{flaherty07}. 
b) Color-magnitude diagram (H$-$K/K) of the sources only detected in H and K bands that have no counterparts 
in our selected GLIMPSE360 catalog.
In both the panels, Class~I and Class~II YSOs are marked by red circles and open blue triangles, respectively. 
In both the panels, the dots in gray color refer the stars with only photospheric emissions. 
The positions of all the identified YSOs are shown in Figure~\ref{fig10a}a.} 
\label{fig9a}
\end{figure*}
\begin{figure*}
\epsscale{0.490}
\plotone{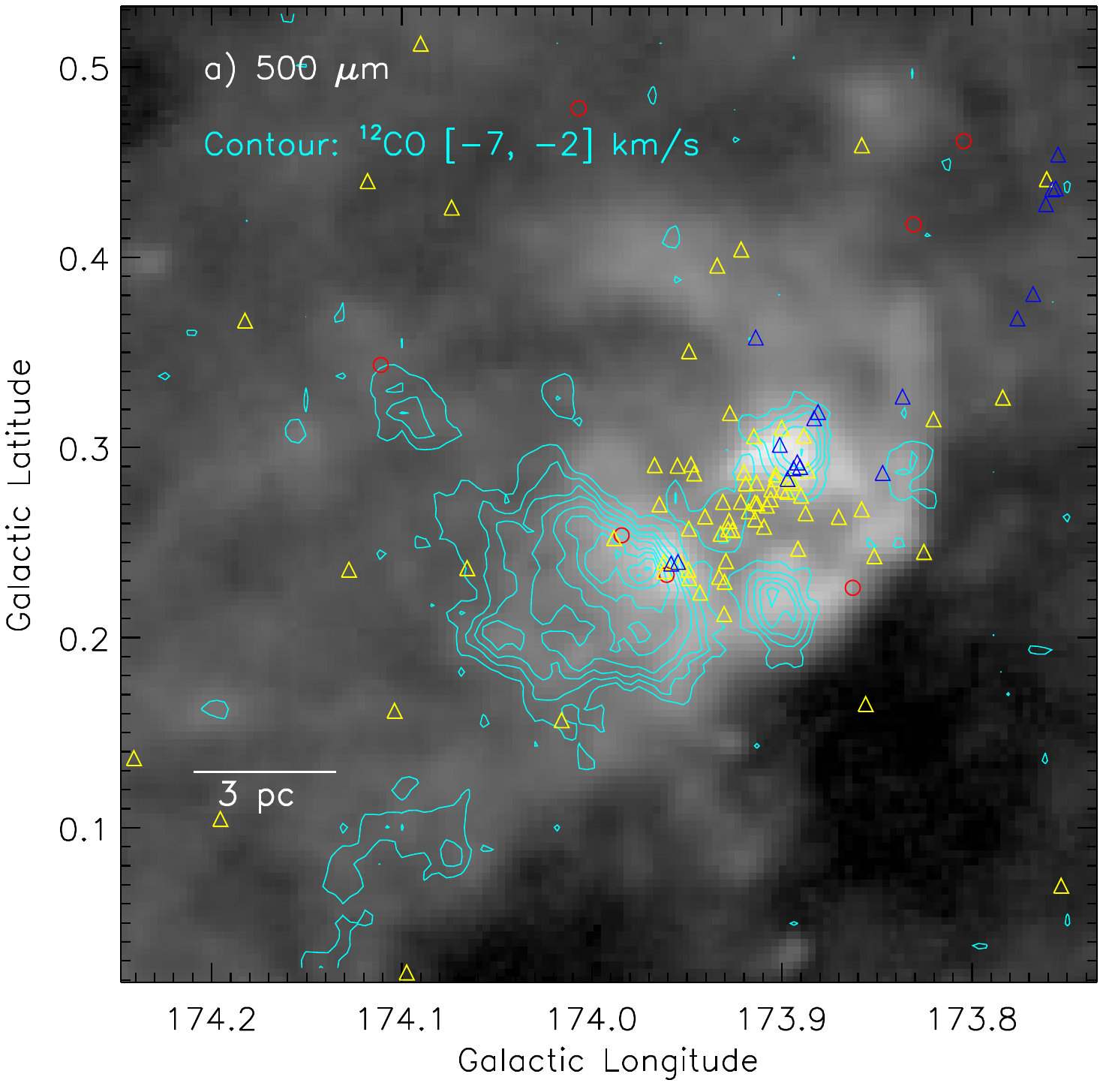}
\epsscale{0.490}
\plotone{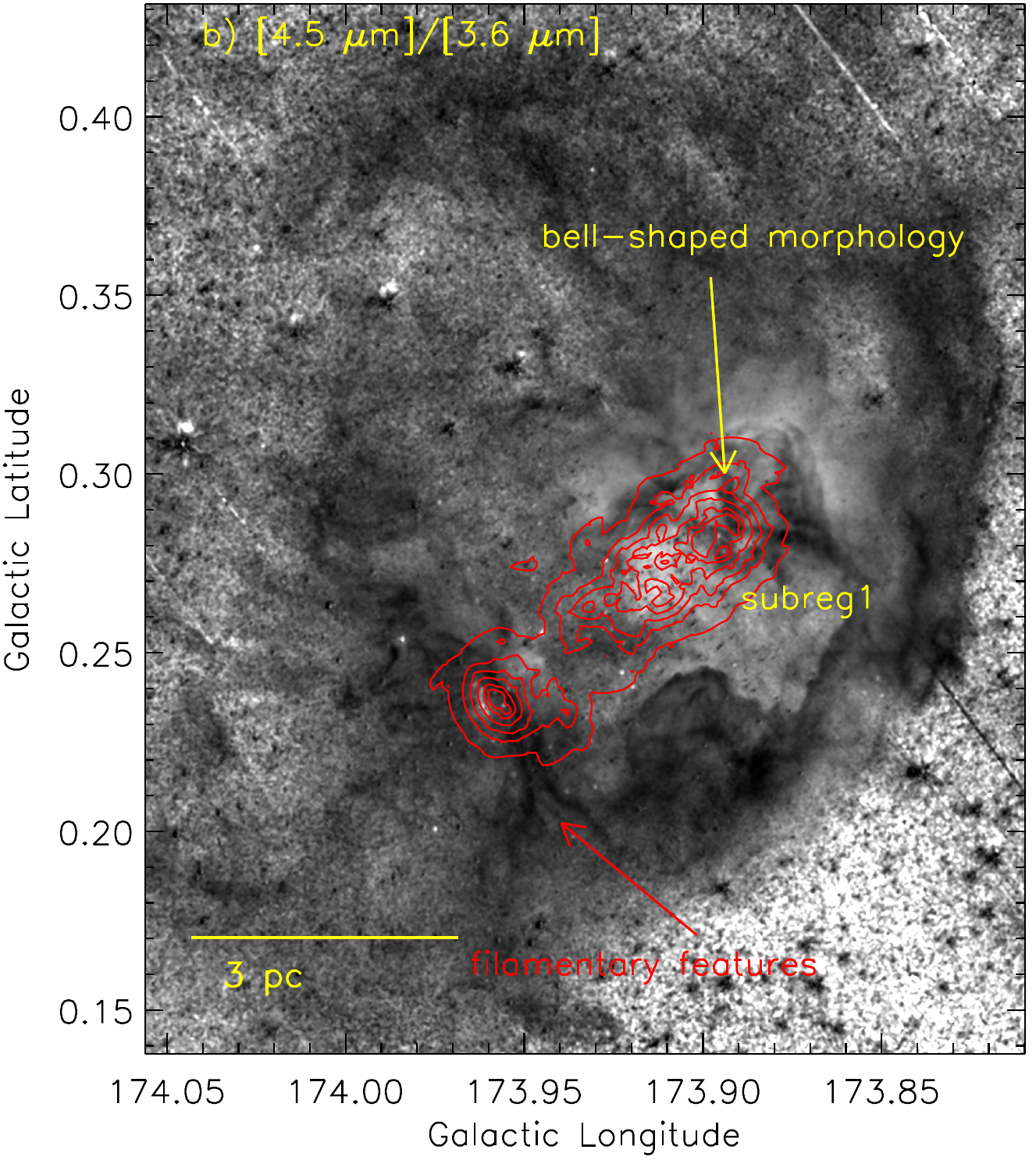}
\caption{\scriptsize The spatial distribution of YSOs identified within the region probed in this paper using the NIR (1--5 $\mu$m) data. 
a) The $^{12}$CO emission and the positions of YSOs are overlaid on the {\it Herschel} 500 $\mu$m map. 
The background map is similar to the one shown in Figure~\ref{fig2a}. 
The positions of Class~I and Class~II YSOs identified within our selected region are shown by circles (in red) 
and triangles (in blue and yellow), respectively. 
The YSOs selected using the H, K, 3.6 $\mu$m, and 4.6 $\mu$m data (see Figure~\ref{fig9a}a) 
are shown by red circles and yellow triangles, whereas the blue triangles represent the YSOs identified using the H and K bands (see Figure~\ref{fig9a}b).
b) The surface density contours (in red) of all the identified YSOs are overlaid on the {\it Spitzer}-IRAC ratio map of 4.5 $\mu$m/3.6 $\mu$m emission (similar area as shown in Figure~\ref{fig3a}a), indicating the star formation activities mainly 
toward the filamentary features/clump1 and the bell-shaped morphology/clump2 (see text for details). 
The contours are shown at 3, 5, 7, 10, 15, and 20 YSOs/pc$^{2}$, from the outer to the inner side. 
In both the panels, the scale bar at the bottom-left corner corresponds to 3 pc (at a distance of 2.3 kpc).} 
\label{fig10a}
\end{figure*}
\begin{figure*}
\epsscale{1}
\plotone{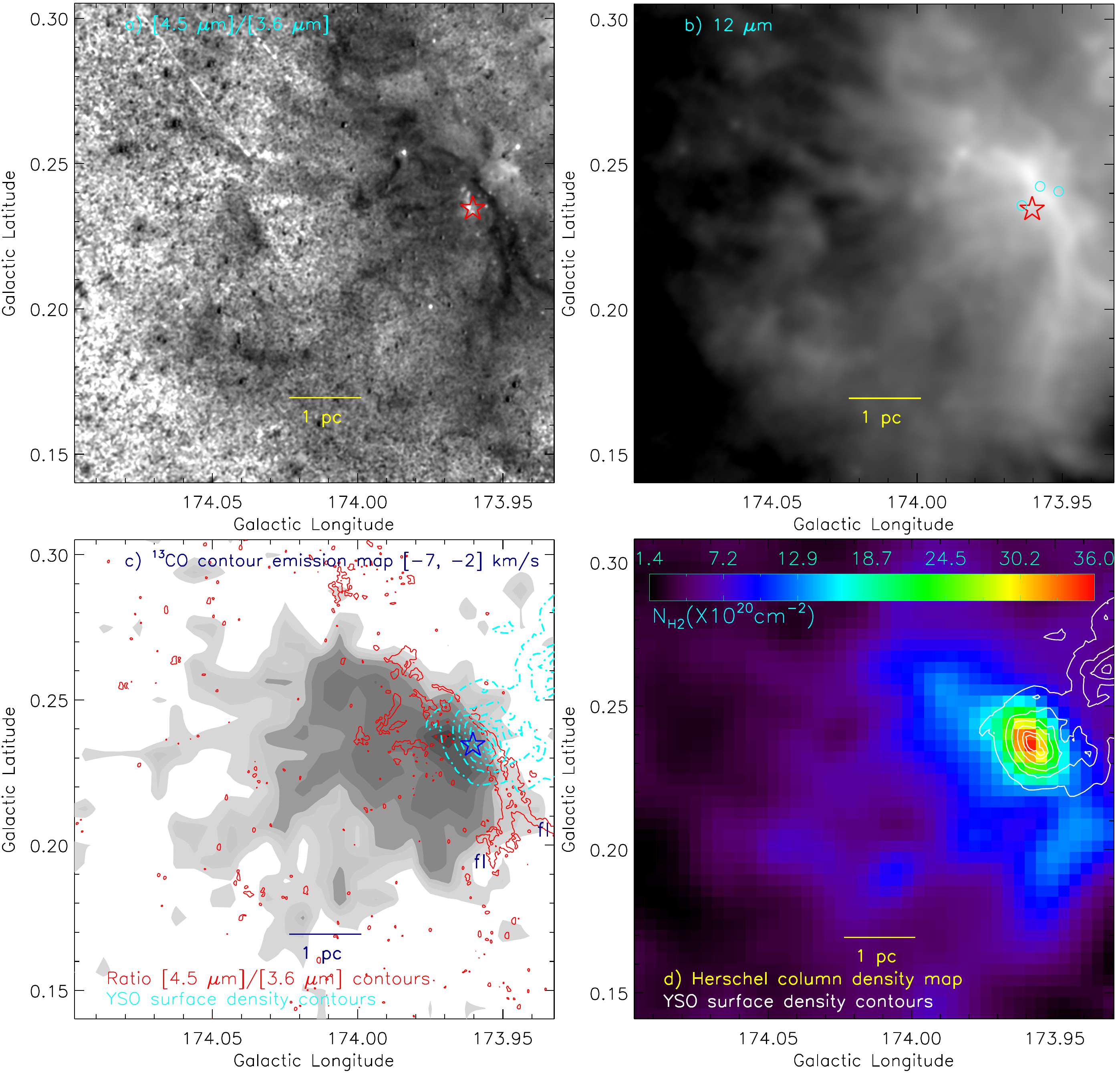}
\caption{\scriptsize Zoomed-in view toward the molecular condensation, conds1, as traced in the integrated $^{12}$CO and $^{13}$CO maps 
(see dotted-dashed box in Figure~\ref{fig6a}). a) {\it Spitzer}-IRAC ratio map of 4.5 $\mu$m/3.6 $\mu$m emission traces the filamentary features. b) {\it WISE} image at 12 $\mu$m also depicts the filamentary features. The positions of three stars having optical polarimetric 
observations \citep[from][]{pandey13} are marked by cyan circles. 
c) The integrated $^{13}$CO(J =1$-$0) emission map of the molecular condensation, conds1. 
The $^{13}$CO contours are shown with levels of 10, 18, 25, 30, 40, 55, 70, 80, 90, and 99\% of the peak value (i.e. 8.251 K km s$^{-1}$). 
The YSO surface density contours (dotted dashed; in cyan color) and filamentary features (in red) are also shown in the map. 
Two filamentary features are highlighted by labels ``fl".
d) {\it Herschel} column density map is overlaid with the YSO surface density contours, revealing a physical connection between 
a cluster of YSOs and a massive clump. 
In the last two panels, the contours are shown at 3, 5, 7, 10, 15, and 20 YSOs/pc$^{2}$, from the outer to the inner side. 
In the first three panels, the position of the {\it Herschel} column density peak is marked by a star symbol. 
Figures exhibit the presence of a cluster of YSOs and a massive clump at the intersection of filamentary features.}
\label{fig11a}
\end{figure*}
\begin{deluxetable}{ccccc}
\tablewidth{0pt} 
\tabletypesize{\scriptsize} 
\tablecaption{Summary of the properties of the {\it Herschel} clumps identified in the S237 region 
(see Figures~\ref{fig5a}b and~\ref{fig5a}c). Column~1 lists the IDs given to the clump. Table also contains 
positions, deconvolved effective radius (R$_{c}$), and clump mass (M$_{clump}$). \label{tab1}} 
\tablehead{ \colhead{ID} & \colhead{{\it l}} & \colhead{{\it b}} & \colhead{R$_{c}$}& \colhead{M$_{clump}$}\\
\colhead{} &  \colhead{[degree]} & \colhead{[degree]} & \colhead{(pc)} &\colhead{($M_\odot$)}}
\startdata 
    1	 &   173.9637  &  0.2325  &	  1.8	&     258.6 \\
    2	  &  173.8937  &  0.2947  &	  1.0	&      84.3  \\
    3	 &   173.9054  &  0.2169  &	  1.0	&      80.2 \\
    4	  &  173.8860  &  0.2130  &	  0.7	&      37.4  \\
    5	  &  173.8704   & 0.2830  &	  0.7	&      35.2 \\
    6	 &   173.8471   & 0.2558  &	  0.7	&      36.0  \\
    7	 &   173.8393  &  0.2830  &	  0.6	&      25.0  \\
    8	  &  174.2360   & 0.3919  &	  0.7	&      27.6   \\
    9	 &   173.9482  &  0.3842  &	  1.3	&      85.7 \\
    10  &    174.0143  &  0.1858 & 	  1.1	&      60.4  \\
    11   &   173.9287  &  0.2480 & 	  0.7	&      24.1 \\
    12  &    174.1582  &  0.3025 & 	  0.5	&      11.7 \\ 
    13  &    173.8393  &  0.3569 & 	  0.6	&      18.5\\
\enddata  
\end{deluxetable}


\begin{thebibliography}{}
%
\bibitem[Andr{\'e} et al.(2010)]{andre10}
Andr{\'e}, P., Men'shchikov, A., Bontemps, S., et al. 2010, A\&A, 518, L102

\bibitem[Arce et al.(2011)]{arce11}
Arce, H.~G., Borkin, M.~A., Goodman, A.~A., Pineda, J.~E.,\& Beaumont, C.~N. 2011, ApJ, 742, 105

\bibitem[Baker \& Burton(1979)]{baker79}
Baker, P.,~L., \& Burton, W.,~B. 1979, A\&AS, 35, 129

\bibitem[Balser et al.(2011)]{balser11}
Balser, D.~S., Rood, R.~T., Bania, T.~M., \& Anderson, L.~D. 2011, ApJ, 738, 27

\bibitem[Bisbas et al.(2009)]{bisbas09}
Bisbas, T.~G., W\"{u}nsch, R., Whitworth, A.~P., \& Hubber, D.~A. 2009, A\&A, 497, 649

\bibitem[Bohlin et al.(1978)]{bohlin78}
Bohlin, R.~C., Savage, B.~D., \& Drake, J.~F. 1978, ApJ, 224, 13233

\bibitem[Bressert et al.(2010)]{bressert10}
Bressert, E., Bastian, N., Gutermuth, R., et al. 2010, MNRAS, 409, 54

\bibitem[Bressert et al.(2012)]{bressert12}
Bressert, E., Ginsburg, A., Bally, J., et al. 2012, ApJ, 758, 28

\bibitem[Brunt(2004)]{brunt04}
Brunt C., 2004, in Clemens D., Shah R., Brainerd T., eds, Proc. of ASP Conference 317. Milky Way Surveys: The
Structure and Evolution of our Galaxy, p. 79

\bibitem[Burton et al.(1978)]{burton78}
Burton, W.,~B., Liszt, H.,~S., \& Baker, P.,~L. 1978, ApJ, 219, 67

\bibitem[Burton \& Liszt(1981)]{burton81}
Burton, W.,~B., \& Liszt, H.,~S. 1981, in Origin of Cosmic Rays, eds. G. Setti, G. Spada, \& A. W. Wolfendale, IAU Symp., 94, 227

\bibitem[Casali et al.(2007)]{casali07}
Casali, M., Adamson, A., Alves de Oliveira, C., et al. 2007, A\&A, 467, 777

\bibitem[Condon et al.(1998)]{condon98}
Condon, J.~J., Cotton, W.~D., Greisen, E.~W., et al. 1998, AJ, 115, 1693

\bibitem[Dale \& Bonnell(2011)]{dale11}
Dale, J.~E.,  \& Bonnell, I.~A. 2011, MNRAS, 414, 321

\bibitem[Davis \& Greenstein(1951)]{davis51}
Davis, L., Jr., \& Greenstein, J. L. 1951, ApJ, 114, 206

\bibitem[Deharveng et al.(2010)]{deharveng10}
Deharveng, L., Schuller, F., Anderson, L.~D., et al. 2010, A\&A, 523, 6

\bibitem[Dewangan \& Anandarao(2011)]{dewangan11}
Dewangan, L.~K., \& Anandarao, B.~G 2011, MNRAS, 414, 1526

\bibitem[Dewangan et al.(2012)]{dewangan12}
Dewangan, L.~K., Ojha, D.~K., Anandarao, B.~G., Ghosh, S.~K., \& Chakraborti, S. 2012, ApJ, 756, 151

\bibitem[Dewangan et al.(2015)]{dewangan15}
Dewangan, L.~K., Luna, A., Ojha, D.~K., et al.  2015, ApJ, 811, 79

\bibitem[Dewangan et al.(2016)]{dewangan16}
Dewangan, L.~K., Ojha, D.~K., Luna, A., et al.  2016, ApJ, 819, 66

\bibitem[Drew et al.(2005)]{drew05}
Drew, J.~E., Greimel, R., Irwin, M.J., et al. 2005, MNRAS, 362, 753

\bibitem[Dyson \& Williams(1980)]{dyson80}
Dyson, J.~E., \& Williams, D.~A. 1980, Physics of the interstellar medium, New York, Halsted Press, 204 p

\bibitem[Elmegreen(1998)]{elmegreen98} 
Elmegreen, B.~G.\ 1998, in ASP Conf. Ser. 148, Origins, ed. C. E. Woodward, J.~M. Shull, \& H.~A. Thronson, Jr. (San Francisco, CA: ASP), 150

\bibitem[Elmegreen(2011)]{elmegreen11} 
Elmegreen, B., G. 2011, EAS Publications Series, EAS Publications Series, 51, 31

\bibitem[Evans et al.(2009)]{evans09}
Evans, N.~J., II, Dunham, M.~M., J\o{}rgensen, J.~K., et al. 2009, ApJS, 181, 321

\bibitem[Flaherty et al.(2007)]{flaherty07}
Flaherty, K.~M., Pipher, J.~L., Megeath, S.~T., et al. 2007, ApJ, 663, 1069

\bibitem[Glushkov et al.(1975)]{glushkov75}
Glushkov, Y.~I., Denisyuk, E.~K.,  \& Karyagina, Z.~V. 1975, A\&A, 39,  481

\bibitem[Griffin et al.(2010)]{griffin10} 
Griffin, M.~J., Abergel, A., Abreu, A, et al. 2010, A\&A, 518L, 3

\bibitem[Gutermuth et al.(2009)]{gutermuth09}
Gutermuth, R.~A., Megeath, S.~T., Myers, P.~C., et al. 2009, ApJS, 184, 18

\bibitem[Heyer et al.(1996)]{heyer96}
Heyer, M.~H., Carpenter, J.~M., \& Ladd, E.~F. 1996, ApJ, 463, 630

\bibitem[Heyer et al.(1998)]{heyer98}
Heyer, M., Brunt, C., Snell, R., et al. 1998, ApJS, 115, 241

\bibitem[Hildebrand(1983)]{hildebrand83} 
Hildebrand, R.~H. 1983, Quarterly Journal of the RAS, 24, 267

\bibitem[Kauffmann et al.(2008)]{kauffmann08}
Kauffmann, J., Bertoldi, F., Bourke, T.~L., Evans, II, N.~J.,\&  Lee, C.~W. 2008, ApJ, 487, 993

\bibitem[Kerton(2005)]{kerton05}
Kerton, C.~R. 2005, ApJ, 623, 235

\bibitem[Krumholz \& McKee(2008)]{krumholz08}
Krumholz, M.,~R., \& McKee, C.,~F. 2008, Nature, 451, 1082

\bibitem[Kwan(1997)]{kwan97} 
Kwan, J. 1997, ApJ, 489, 284

\bibitem[Lawrence et al.(2007)]{lawrence07}
Lawrence, A., Warren, S.~J., Almaini, O., et al. 2007, MNRAS, 379, 1599

\bibitem[Leisawitz et al.(1989)]{leisawitz89}
Leisawitz, D., Bash, F.~N., \& Thaddeus, P. 1989, ApJS, 70, 731

\bibitem[Lim et al.(2015)]{lim15}
Lim, B., Sung, H., Hur, H., et al. 2015, JKAS, 48, 343

\bibitem[Liszt et al.(1981)]{liszt81}
Liszt, H.,~S., Burton, W.,~B., \& Bania, T.,~M. 1981, ApJ, 246, 74

\bibitem[Lucas et al.(2008)]{lucas08}
Lucas, P.~W., Hoare, M.~G., Longmore, A., et al. 2008, MNRAS, 391, 1281

\bibitem[Mallick et al.(2015)]{mallick15}
Mallick, K.~K., Ojha, D.~K., Tamura, M., et al. 2015, MNRAS, 447, 2307

\bibitem[Matsakis et al.(1976)]{matsakis76}
Matsakis, D.~N., Evans, N.~J., II, Sato, T., \& Zuckerman, B. 1976, AJ, 81, 172

\bibitem[Molinari et al.(2010)]{molinari10}
Molinari, S., Swinyard, B., Bally, J., et al., 2010, A\&A, 518, L100

\bibitem[Myers (2009)]{myers09} 
Myers, P.~C. 2009, ApJ, 700, 1609

\bibitem[Nakamura et al.(2014)]{nakamura14}
Nakamura, F., Sugitani, K., Tanaka, T., et al 2014, ApJL, 791, L23

\bibitem[Oskinova et al.(2011)]{oskinova11} 
Oskinova, L.~M., Todt, H., Ignace, R., et al.\ 2011, MNRAS, 416, 1456 

\bibitem[Ott(2010)]{ott10}
Ott, S. 2010, in Astronomical Society of the Pacic Conference
Series, Vol. 434, Astronomical Data Analysis Software and
Systems XIX, ed. Y. Mizumoto, K.-I. Morita, \& M. Ohishi, 139

\bibitem[Panagia(1973)]{panagia73} 
Panagia, N. 1973, AJ, 78, 929 

\bibitem[Pandey et al.(2013)]{pandey13} 	
Pandey, A.~K., Eswaraiah, C., Sharma, S. et al. 2013, ApJ, 764, 172

\bibitem[Poglitsch et al.(2010)]{poglitsch10}	
Poglitsch, A., Waelkens, C., Geis, N., et al. 2010, A\&A, 518L, 2

\bibitem[Povich et al.(2007)]{povich07} 
Povich, M.~S., Stone, J.~M., Churchwell, E., et al. 2007, ApJ, 660, 346

\bibitem[Reipurth(2008)]{reipurth08} 	
Reipurth, B., 2008, Handbook of Star Forming Regions Vol. I, Astronomical Society of the Pacific, Ed. B. Reipurth

\bibitem[Schneider et al.(2012)]{schneider12}
Schneider, N., Csengeri, T., Hennemann, M., et al. 2012, A\&A, 540, L11

\bibitem[Skrutskie et al.(2006)]{skrutskie06}
Skrutskie, M.~F., Cutri, R.~M., Stiening, R., et al. 2006, AJ, 131, 1163
%

\bibitem[Sugitani et al.(2011)]{sugitani11}
Sugitani, K., Nakamura, F., Watanabe, M., et al. 2011, ApJ, 734, 63

\bibitem[Tan et al.(2014)]{tan14} 	
Tan, J.~C., Beltr\'an, M.~T., et al. 2014, aXriv 1402.0919

\bibitem[Taylor et al.(2003)]{taylor03}
Taylor, A.,~R., Gibson, S.,~J., Peracaula, M., et al. 2003, AJ, 125, 3145

\bibitem[Voges et al.(1999)]{voges99}
Voges, W., Aschenbach, B., Boller, Th., et al. 1999, A\&A, 349, 389

\bibitem[Voges et al.(2000)]{voges00}
Voges, W., Aschenbach, B., Boller, Th., et al. 2000, IAU Circ. 7432

\bibitem[Watson et al.(2008)]{watson08}
Watson, C., Povich, M.~S., Churchwell, E.~B., et al. 2008, ApJ, 681, 1341

\bibitem[Watson et al.(2010)]{watson10}
Watson, C., Hanspal, U., \& Mensistu, A. 2010, ApJ, 716, 1478

\bibitem[Whitney et al.(2011)]{whitney11}
Whitney, B., Benjamin, R., Meade, M., et al. 2011, Bulletin of the American Astronomical Society, Vol. 43

\bibitem[Williams et al.(1994)]{williams94} 
Williams, J. P., de Geus, E. J., \& Blitz, L. 1994, ApJ, 428, 693

\bibitem[Wright et al.(2010)]{wright10}
Wright, E.~L., Eisenhardt, P.~R.~M., Mainzer, et al. 2010, AJ, 140, 1868

\bibitem[Yan et al.(2016)]{yan16}
Yan, Q.~Z., Xu, Y., Zhang, B., et al. 2016, arXiv:1609.03051

\bibitem[Yang et al.(2002)]{yang02}
Yang, J., Jiang, Z., Wang, M., Ju, B., \& Wang, H. 2002, ApJS, 141, 157

\bibitem[Zinnecker \& Yorke(2007)]{zin07} 
Zinnecker, H., \& Yorke, H.~W. 2007, ARA\&A, 45, 481 
\end{thebibliography}
\end{document}